\documentclass[twocolumn,astrosymb]{aastex631}
\shorttitle{AGN Feedback in Quiescent Galaxies at Cosmic Noon}
\shortauthors{Bugiani et al.}
\usepackage{amsmath,amsfonts,amssymb,mathtools}
\usepackage{amsthm}
\usepackage{graphicx}

\usepackage{appendix}
\usepackage{tabularx}
\usepackage{float}
\usepackage{hyperref}

\begin{document}

\title{AGN Feedback in Quiescent Galaxies at Cosmic Noon Traced by Ionized Gas Emission}

\author{Letizia Bugiani}
\affiliation{Dipartimento di Fisica e Astronomia, Università di Bologna, Bologna, Italy.}
\affiliation{INAF, Osservatorio di Astrofisica e Scienza dello Spazio, Via Piero Gobetti 93/3, I-40129, Bologna, Italy.}

\author{Sirio Belli}
\affiliation{Dipartimento di Fisica e Astronomia, Università di Bologna, Bologna, Italy.}

\author{Minjung Park}
\affiliation{Center for Astrophysics | Harvard \& Smithsonian, Cambridge, MA, USA.}

\author{Rebecca L. Davies}
\affiliation{Centre for Astrophysics and Supercomputing, Swinburne University of Technology, Hawthorn, Victoria, Australia.}
\affiliation{ARC Centre of Excellence for All Sky Astrophysics in 3 Dimensions (ASTRO 3D), Australia.}

\author{J. Trevor Mendel}
\affiliation{ARC Centre of Excellence for All Sky Astrophysics in 3 Dimensions (ASTRO 3D), Australia.}
\affiliation{Research School of Astronomy and Astrophysics, Australian National University, Canberra, ACT, Australia.}

\author{Benjamin D. Johnson}
\affiliation{Center for Astrophysics | Harvard \& Smithsonian, Cambridge, MA, USA.}

\author{Amir H. Khoram}
\affiliation{Dipartimento di Fisica e Astronomia, Università di Bologna, Bologna, Italy.}
\affiliation{INAF, Osservatorio di Astrofisica e Scienza dello Spazio, Via Piero Gobetti 93/3, I-40129, Bologna, Italy.}

\author{Chlo\"{e}  Benton}
\affiliation{Department for Astrophysical and Planetary Science, University of Colorado, Boulder, CO, USA.}

\author{Andrea Cimatti}
\affiliation{Dipartimento di Fisica e Astronomia, Università di Bologna, Bologna, Italy.}
\affiliation{INAF, Osservatorio di Astrofisica e Scienza dello Spazio, Via Piero Gobetti 93/3, I-40129, Bologna, Italy.}

\author{Charlie Conroy}
\affiliation{Center for Astrophysics | Harvard \& Smithsonian, Cambridge, MA, USA.}

\author{Razieh Emami}
\affiliation{Center for Astrophysics | Harvard \& Smithsonian, Cambridge, MA, USA.}

\author{Joel Leja}
\affiliation{Department of Astronomy \& Astrophysics, The Pennsylvania State University, University Park, PA, USA.}
\affiliation{Institute for Gravitation and the Cosmos, The Pennsylvania State University, University Park, PA, USA.}
\affiliation{Institute for Computational \& Data Sciences, The Pennsylvania State University, University Park, PA, USA.}

\author{Yijia Li}
\affiliation{Department of Astronomy \& Astrophysics, The Pennsylvania State University, University Park, PA, USA.}
\affiliation{Institute for Gravitation and the Cosmos, The Pennsylvania State University, University Park, PA, USA.}

\author{Gabriel Maheson}
\affiliation{Kavli Institute for Cosmology, University of Cambridge, Cambridge, UK.}
\affiliation{Cavendish Laboratory, University of Cambridge, Cambridge, UK.}

\author{Elijah P. Mathews}
\affiliation{Department of Astronomy \& Astrophysics, The Pennsylvania State University, University Park, PA, USA.}
\affiliation{Institute for Gravitation and the Cosmos, The Pennsylvania State University, University Park, PA, USA.}
\affiliation{Institute for Computational \& Data Sciences, The Pennsylvania State University, University Park, PA, USA.}

\author{Rohan P. Naidu}
\affiliation{MIT Kavli Institute for Astrophysics and Space Research, Cambridge, MA, USA.}

\author{Erica J. Nelson}
\affiliation{Department for Astrophysical and Planetary Science, University of Colorado, Boulder, CO, USA.}

\author{Sandro Tacchella}
\affiliation{Kavli Institute for Cosmology, University of Cambridge, Cambridge, UK.}
\affiliation{Cavendish Laboratory, University of Cambridge, Cambridge, UK.}

\author{Bryan A. Terrazas}
\affiliation{Columbia Astrophysics Laboratory, Columbia University, New York, NY, USA.}

\author{Rainer Weinberger}
\affiliation{Leibniz Institute for Astrophysics, Potsdam, Germany.}

%\collaboration{20}{(AAS Journals Data Editors)}



\begin{abstract}

We analyze ionized gas emission lines in deep rest-frame optical spectra of 14 massive ($\log M_*>10.2$ M$_{\odot}$) quiescent galaxies at redshift \(1.7<z<3.5\) observed with JWST/NIRSpec by the Blue Jay survey. Robust detection of emission lines in 71\% of the sample indicates the presence of ongoing ionizing sources in this passive population. The H$\alpha$ line luminosities confirm that the population is quiescent, with star formation rates that are at least ten times lower than the main sequence of star formation at $z\sim2$.
The quiescent sample is clearly separate from the star-forming population in line diagnostic diagrams, and occupies a region usually populated by active galactic nuclei (AGN). Analysis of the observed line ratios, equivalent widths, and velocity dispersions leads us to conclude that in most cases the gas is ionized by AGN activity, despite the lack of X-ray detections. We measure generally low value of bolometric luminosity $L_{BOL} \sim 10^{44}$ erg s$^{-1}$ and of Eddington ratios $\lambda \sim 10^{-2} - 10^{-3}$ for the central engines, typical of low-luminosity AGN. A subset of the sample also hosts ionized and neutral outflows, with ionized outflows velocities of the order of $\sim1000$ km s$^{-1}$. Our results show, for the first time using a representative sample, that low luminosity AGN are extremely common among quiescent galaxies at high redshift. These low luminosity AGN may play a key role in quenching star formation and in maintaining massive galaxies quiescent from Cosmic Noon to $z\sim0$.
\end{abstract}

%\keywords{Classical Novae (251) $-$- Ultraviolet astronomy(1736) $-$- History of astronomy(1868) $-$- Interdisciplinary astronomy(804)}


\section{Introduction} \label{sec:intro}

At Cosmic Noon, the $z\sim2$ peak of the cosmic star formation history \citep{MADAU2014}, most galaxies host a large amount of gas. However, this is also the epoch when the population of massive galaxies begins to transform into quiescent, gas-poor systems. The quenching of star formation is one of the key moments in a galaxy's lifetime, but the physics behind it is still poorly understood, and requires more detailed studies of the first generation of quiescent galaxies.

Rest-frame optical emission lines due to ionized gas can provide a wealth of information about the physical conditions of high-redshift galaxies \citep[][and references therein]{emission_lines_bible}. 
In star-forming galaxies, the ionization of the interstellar medium (ISM) is caused by young, massive stars; strong emission lines can also be produced by Active Galactic Nuclei (AGN), where the ionizing photon field is due to the actively feeding supermassive black hole (SMBH) in the center of the galaxy. Gas in the ISM can also be excited by shocks, which are more difficult to detect and can be caused by several different phenomena, including galactic-scale outflows, galaxy interactions, ram pressure stripping, and AGN-related activity such as jets and winds \citep{emission_lines_bible}. Mergers can also produce widespread shocks throughout galaxies which significantly affect the emission-line spectrum \citep{MEDLING15}, and stellar wind-induced shocks have been observed both in star-forming and starburst galaxies \citep{Heckman17,Ho16}.

Observations of ionized gas in high-redshift quiescent galaxies are necessary in order to constrain the physical conditions of these systems and to develop a comprehensive picture of galaxy quenching. However, the small amount of ionized gas left in these galaxies makes it challenging to detect and measure the emission lines, particularly at high redshift. Detections of very weak optical emission lines can be obtained more easily in the local universe.
The presence of warm ($\sim 10^4$K) ionized gas in local quiescent galaxies, estimated to represent around 1\% of their total ISM, has been known for a long time \citep{buson93,phillips86}: proposed explanations for the presence of this ionized phase have included external accretion of warm ISM from a companion galaxy \citep{bertola92,kim98}, heat transfer from hot ($\sim10^7 $K) phase gas in the halo \citep{sparks98,voigtdoanhue90}, low-power nuclear activity (as in LINERs -- Low-ionization nuclear emission-line regions);  \citep{kim98,shields92} and shocks \citep{heckman89}. The prevailing theory attributes the presence of warm ($\sim 10^4$K) gas to photoionization by hot evolved low-mass stars (HOLMES); \citep{binette94,macchetto96}. More recently, \cite{CidFernades11,CideFernandes10} used emission line ratios to distinguish so-called Emission Line Retired Galaxies (EL-RGs) with persistent ionized gas due to HOLMES photoionizing flux from true LINERs, i.e. galaxies that host a low-power AGN \citep[see also][]{belfiore16}.

Beyond the limits of the local Universe the picture becomes much less clear, as the observations becomes more challenging; at $z>1$ it is particularly difficult to detect faint optical emission lines from ionized gas because the rest-frame optical spectrum is observed at near-infrared wavelengths. Deep exposures with the largest ground-based telescopes have led to the detection of [NII] and H$\alpha$ emission lines in individual quiescent galaxies \citep{Belli17_kmos, heavymetalsurvey} or in stacked spectra \citep{Belli19}. The faintness of the observed emission lines rules out a substantial star formation rate in these galaxies; moreover, the [NII]/H$\alpha$ flux ratio is typically elevated and not consistent with ionization from young stars in HII regions.
More detailed measurements of faint emission lines are possible in those rare $z\sim2$ quiescent galaxies that are gravitationally lensed by a foreground cluster \citep{newman15, newman18, toft17, man21}. These studies confirm that young stars in HII regions contribute very little to both the ionization of the gas and the overall growth of the galaxies (i.e., the measured specific star formation rates are very low). The ionization mechanism is often attributed to AGN feedback, but very little direct evidence for this exists given the scarcity of data; for example, the four emission lines required for the standard BPT diagram \citep{BPT_original} have been measured in just two lensed galaxies \citep{newman18}.

Recently, the launch of JWST has begun a new era of sensitive near-infrared spectroscopy, free from contamination by atmospheric emission and absorption. Early JWST spectra of individual quiescent galaxies at $z>2$ have revealed the presence of emission lines clearly due to AGN activity \citep{Carnall23, 11142_article, deugenio23}, showcasing the potential of space-based spectroscopy for the study of faint emission from ionized gas at high redshift. 
In this work, we present the first comprehensive study of rest-frame optical emission lines in a representative sample of massive quiescent galaxies at Cosmic Noon, based on deep spectroscopy obtained by the Blue Jay survey using the NIRSpec instrument onboard JWST. 
In Section~\ref{sec:data} we describe the spectroscopic data and the selection of the quiescent sample. In Section~\ref{sec:fitting} we present the spectral fitting of the emission lines: the results derived from these fits are explored in Section~\ref{sec:results_1}, where we investigate the origin of the observed ionized gas using several line diagnostics, and in Section~\ref{sfr}, where we derive upper limits on the current star formation rates. In Section~\ref{subsec:outflows} we conduct a more in-depth spectral analysis on a subset of galaxies which show signs of powerful ionized gas outflows and derive the outflow velocities. In Section~\ref{sec:disc} we gather all the evidence obtained from our data analysis to try and identify the main ionization mechanisms for this sample of \(z\sim2\) quiescent galaxies, and put our results in context within the overall picture of quenching in high-redshift systems. Throughout the paper, we adopt the WMAP9 cosmology \citep{hinshaw13}. 

\section{Data and sample selection} \label{sec:data}

This work is based on spectroscopic data obtained by the Blue Jay survey. In this section, we first provide an overview of the Blue-Jay survey, followed by the spectral modeling that led to the selection of the sample of quiescent galaxies presented in this paper. 

\begin{figure*}[t]
    \includegraphics[width=\textwidth]{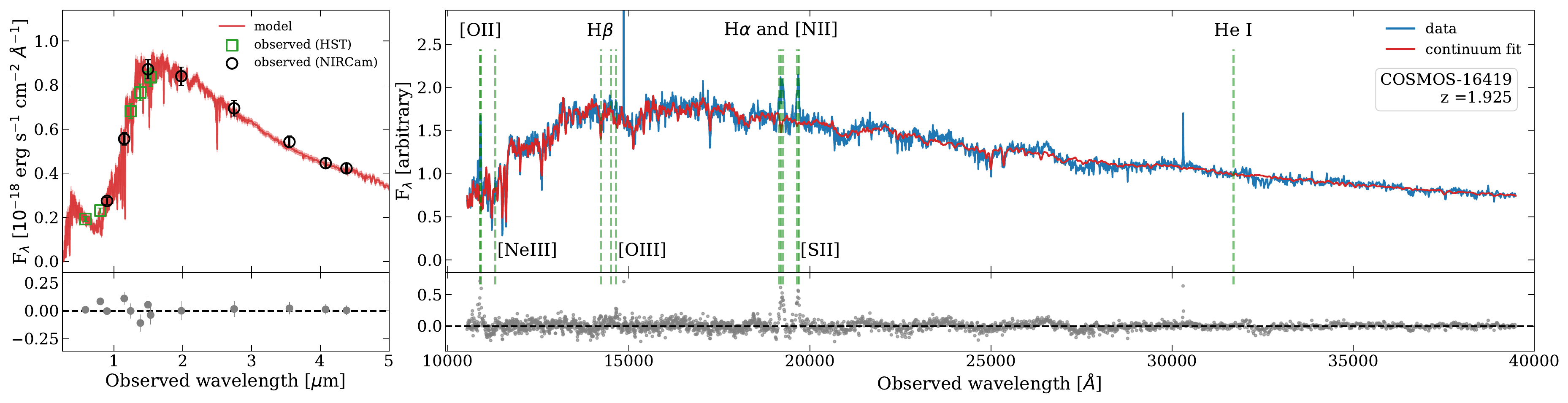}
    \caption{Example of stellar population fit for a quiescent galaxy in the sample. Left panel: photometric data (black circles from NIRCam, green squares from HST) with the best-fit Prospector model (red line). Right panel: observed NIRSpec spectrum (blue) overlaid with the best fit from Prospector (red), adopting a wide wavelength range. Highlighted emission lines (green dashed lines) have been masked during the stellar continuum fit. Residuals of the fits are plotted in grey under each panel.}
    \label{fig:example_spectrum}
\end{figure*}

\subsection{The Blue Jay survey}
The Blue Jay survey is a medium-sized JWST Cycle-1 program (GO 1810; PI: S. Belli). Observations targeted 147 galaxies at Cosmic Noon and 4 filler galaxies at $z\sim6$, for a total of 151 targets, and were performed using the NIRSpec instrument in multi-object mode with the Micro-Shutter Assembly (MSA). The targets were selected in the COSMOS field using Hubble Space Telescope (HST) observations by the Cosmic Assembly Near-Infrared Deep Extragalactic Legacy Survey \cite[CANDELS;][]{grogin11,Koekemoer11} and the photometric catalog released by the 3D-HST team \citep{BRAMMER12,Skelton14,Momcheva16}, which includes both ground- and space-based data. Photometric redshifts and stellar mass estimates of the parent catalog were obtained through SED fitting, then a catalog of target galaxies was selected to ensure a roughly uniform coverage in both redshift and stellar mass, yielding a representative sample of galaxies with stellar mass \(9 \lesssim \log(M_*/M_{\odot}) \lesssim 11.5\) in the redshift range $1.7<z<3.5$.

Each target was observed through an MSA slitlet composed of at least two shutters. Empty shutters were used to construct a master background spectrum, which was then subtracted by each target spectrum. Targets were observed using the G140M/F100LP, G235M/F170LP and G395M/F290LP medium-resolution gratings (R $\sim$ 1000) with exposure times of 13h, 3.2h and 1.6h respectively. The individual grating spectra were combined to produce spectra with a continuous wavelength coverage from 1 to 5 $\mu$m, but with occasional gaps due to the space between the two NIRSpec detectors.
The full description of the Blue Jay survey design and data reduction process will be provided in the survey paper (Belli et al, in prep.).

\subsection{Stellar continuum fit and flux calibration \label{sec:slit_loss_flux_calibration}}
In order to measure the emission line fluxes from the observed spectra, we need to account for two issues. First, given the small size of the MSA shutters, the NIRSpec spectra probe only a fraction of the light emitted by the target and are therefore affected by heavy slit losses: any line flux measured from the spectra will therefore underestimate the true line flux. Second, the spectra of massive galaxies typically have a strong stellar continuum, which must be subtracted from the observed spectrum to obtain the spectrum of the ionized gas. We solve both problems by fitting stellar population models to the combined photometric and spectroscopic data. 

The stellar continuum fitting procedure and the analysis of the stellar population properties, including the star formation histories, are detailed in \cite{MJ2024}; here we only give a summary of the methodology.
The fits are carried out using the fully Bayesian code \texttt{Prospector} \citep{Prospector_article21}, following the approach outlined in \cite{Park23} and \cite{Tacchella22}. The \texttt{Prospector} code adopts the stellar population synthesis model FSPS \citep{conroy09} and a set of free parameters to generate a synthetic galactic spectral energy distribution which can then be fitted to the observations. The model is highly flexible and includes a non-parametric star formation history fitted over 14 age bins, dust attenuation, and dust emission at longer wavelengths. 
Prospector models can be fitted both to photometric and spectral data: in addition to the NIRSpec spectra, we make use of HST and JWST photometry from the 3D-HST survey \citep[5 HST/ACS+WFC3 filters,][]{Skelton14} and the Primer survey \citep[8 NIRCam filters,][]{PRIMER}. The complete photometric catalog of the Blue Jay survey galaxies is publicly available \footnote{\hyperref[]{https://zenodo.org/records/13292819}} and will be presented in detail in a future paper. 

We fit exclusively photometric data for a minority of galaxies where spectral features are not identified or whose SED cannot be reliably modeled, such as in spectra affected by undersampling issues, galaxies undergoing mergers or containing broad AGN lines. For the rest of the sample (113 galaxies), a single \texttt{Prospector} model is fitted both to the NIRSpec spectrum and the broadband photometry. All the emission lines analyzed in this work are marginalized over during the fit, to avoid potential contamination on the estimated star-formation rates due to AGN emission lines. A polynomial distortion is applied to the spectrum in order to match it with the multi-band photometry. This method ties the flux calibration of the spectrum to that of the broadband photometry, which is far more robust; and it also accounts for slit losses if we assume that the emission probed by the MSA shutters is a scaled down version of the total emission coming from the whole galaxy, i.e., assuming spatially uniform emission across the target. We also assume that the ionized gas emission has the same distribution of the stellar emission, so that the same flux calibration can be applied to both components. We note that this assumption does not significantly impact our results because they are mostly based on line ratios and equivalent widths. 

For each galaxy, the Blue Jay spectra cover a wide wavelength range due to the use of three separate gratings. However, the default \texttt{Prospector} fits use only the rest-frame 4000-6700 \r A region, which is easier to model with synthetic stellar populations and is less prone to systematic uncertainties, since the calibration polynomial can artificially change the strength of the 4000-\r A continuum break. This choice leads to the most robust measurements of the galaxy physical properties (see \citealt{MJ2024}). On the other hand, several emission lines from ionized gas lie outside this spectral region, and so \texttt{Prospector} was run again on a wider wavelength range, from 3700 to 13700 \r A; see Figure~\ref{fig:example_spectrum} for an example. Both versions of the \texttt{Prospector} fit use the same set of broadband photometry and adopt the same model.
In the present work, we take the physical measurements such as stellar mass and star formation rate from the default fits with a narrow wavelength range, but we adopt the fits with extended wavelength range when measuring the emission lines.

\subsection{Quiescent sample selection}
\begin{figure}
    \includegraphics[width=\linewidth]{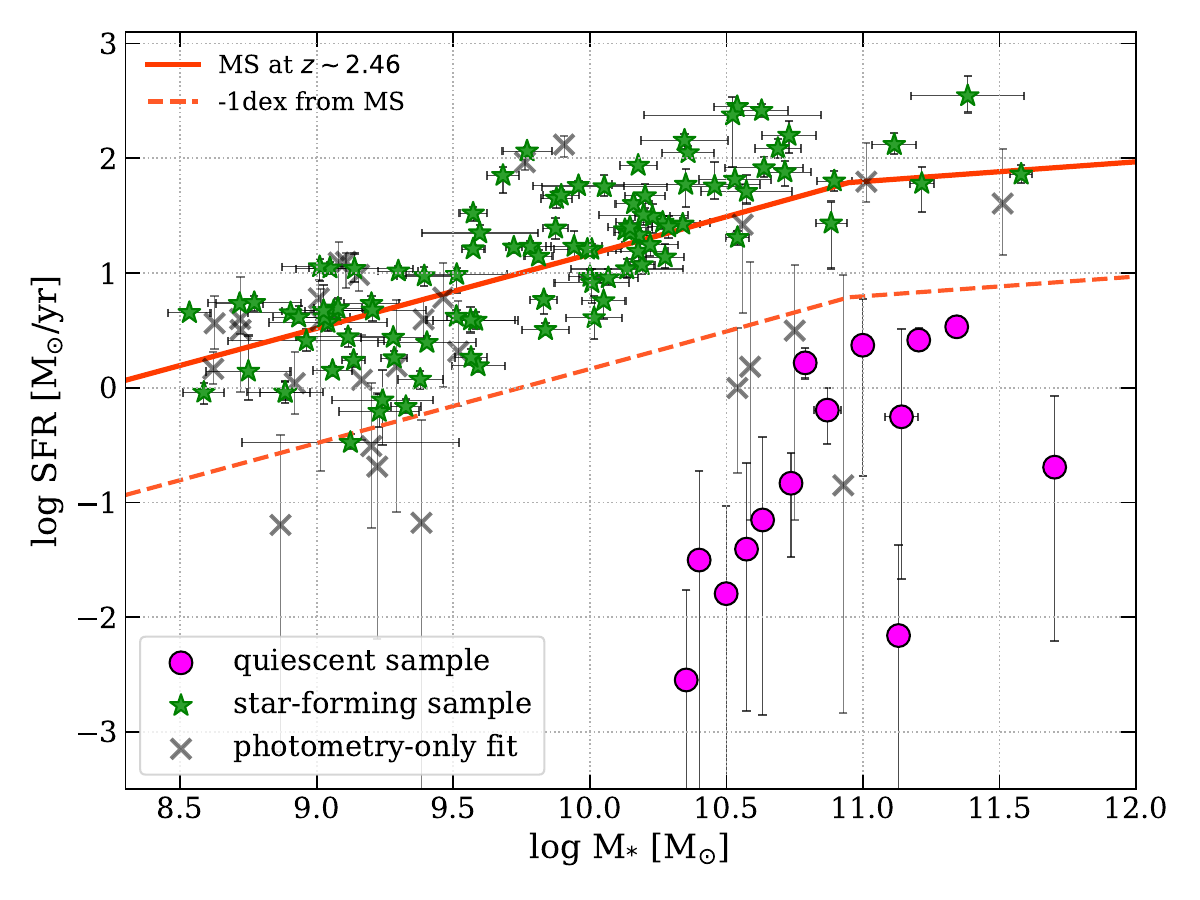}
    \caption{Stellar mass vs SFR (averaged over the last 30 Myr) for the Blue Jay sample. The orange solid line marks the star-formation main sequence at $z=2.46$ from \cite{leja2022}; the orange shaded region shows the $\pm$ 1 dex limits.}
    \label{fig:second_selection}
\end{figure}

To select the quiescent galaxies from the parent sample, we adopt the SFR-based selection of \cite{MJ2024}. Using the star formation history derived with Prospector, we take the mean SFR over the last 30 Myr as the ``instantaneous'' measurement, and compare it to the main sequence of star-forming galaxies measured by \cite{leja2022} at each galaxy's redshift. We then classify as quiescent those galaxies whose 84th percentile of the SFR posterior distribution lies more that 1 dex below the main sequence. Galaxies with photometry-only Prospector fits are automatically excluded from this selection.  
We thus obtain a sample of 14 quiescent galaxies, represented by the magenta circles in Figure~\ref{fig:second_selection}, which illustrates the SFR vs stellar mass diagram of the whole Blue Jay sample, as well as the  \cite{leja2022} star-formation main sequence at $z\sim2.46$ (median redshift of the Blue Jay sample) plotted for visual reference. The main properties of the quiescent sample are listed in Table~\ref{prospector}; we note that all the selected galaxies are more massive than $10^{10.2} ~\text{M}_{\odot}$. Lastly, we have to consider whether contamination from possible AGN continuum emission may be affecting the broadband photometry and biasing our selection. All SEDs in our sample follow a similar shape to the example one that we have included in the left panel of Fig.~\ref{fig:example_spectrum}: this has the typical rest-frame optical profile of a quiescent galaxy with old and red stellar populations, in which the 4000 $\AA$/ Balmer break is clearly visible along with other deep stellar absorption lines. In the observed photometry points, we can see that this break — only due to the stellar component of the galaxy — is very steep and brings a drop of more than 80\% in flux. Considering an additional component due to the AGN continuum may be present, this must be lower than 20\% of the total flux coming from the galaxy in the optical band, which is comparable to the errors on Prospector-estimated quantities such as the SFR. Moreover, the AGN continuum emission could be affected by dust and thus be more relevant in the mid-IR band, but this would not affect the results of this work since that region lies outside of the photometric range analyzed in this work. Thus, we conclude that relevant contamination from AGN continuum emission is not an issue affecting the selected quiescent galaxies in our sample.

\begin{deluxetable*}{cccccccccc}
\tablecaption{Overview of selected quiescent galaxies properties \label{prospector}}

\tablehead{\colhead{COSMOS ID} & \colhead{R.A.} & \colhead{Decl.} & \colhead{} & \multicolumn{2}{c}{Rest-frame colors} & \colhead{} & \multicolumn{3}{c}{\texttt{Prospector} fitted parameters} \\
\cline{5-6} \cline{8-10}
\colhead{} & \colhead{(hh:mm:ss)} & \colhead{(dd:mm:ss)} & \colhead{} & \colhead{$U-V$} & \colhead{$V-J$} & \colhead{} & \colhead{z}  & \colhead{log M$_*$/M$_{\odot}$} & \colhead{SFR (M$_{\odot}$/yr)} }

\startdata
7549 & 10:00:28.83 & 2:15:20.06 & & 2.0 & 1.3 & & 2.627 & $10.79_{-0.03}^{+0.03}$ & $1.6_{-0.4}^{+0.6}$ \\
8013 & 10:00:21.32 & 2:15:41.77 & & 1.5 & 0.7 & & 1.689 & $10.63_{-0.03}^{+0.02}$ & $0.07_{-0.07}^{+0.30}$ \\
8469 & 10:00:21.21 & 2:15:56.53 & & 1.5 & 1.0 & & 1.868 & $10.50_{-0.04}^{+0.08}$ & $0.02_{-0.01}^{+0.08}$ \\
9395 & 10:00:30.16 & 2:16:30.90 & & 1.6 & 0.7 & & 2.127 & $10.74_{-0.02}^{+0.03}$ & $0.1_{-0.1}^{+0.1}$ \\
10128 & 10:00:22.20 & 2:17:01.57 & & 2.0 & 1.2 & & 1.852 & $11.20_{-0.03}^{+0.05}$ & $2.6_{-0.4}^{+0.7}$ \\
10339 & 10:00:22.53 & 2:17:05.00 & & 1.7 & 0.7 & & 2.363 & $10.40_{-0.03}^{+0.03}$ & $0.03_{-0.03}^{+0.16}$ \\
10400 & 10:00:20.41 & 2:17:07.49 & & 2.2 & 1.1 & & 2.098 & $10.35_{-0.04}^{+0.04}$ & $0.003_{-0.003}^{+0.014}$ \\
10565 & 10:00:22.61 & 2:17:14.16 & & 1.8 & 1.0 & & 2.441 & $10.87_{-0.05}^{+0.05}$ & $0.6_{-0.3}^{+0.4}$ \\
10592 & 10:00:18.97 & 2:17:17.67 & & 1.8 & 1.0 & & 1.801 & $11.13_{-0.02}^{+0.03}$ & $0.007_{-0.007}^{+0.035}$ \\
11142 & 10:00:17.59 & 2:17:35.84 & & 1.7 & 1.0 & & 2.444 & $10.89_{-0.02}^{+0.04}$ & $2_{-2}^{+4}$ \\
11494 & 10:00:17.73 & 2:17:52.72 & & 1.9 & 1.0 & & 2.091 & $11.70_{-0.04}^{+0.04}$ & $0.2_{-0.2}^{+0.6}$ \\
16419 & 10:00:22.95 & 2:21:00.25 & & 1.6 & 1.0 & & 1.925 & $11.34_{-0.02}^{+0.02}$ & $3.4_{-0.5}^{+0.6}$ \\
18668 & 10:00:31.03 & 2:22:10.43 & & 1.9 & 1.5 & & 2.086 & $11.14_{-0.06}^{+0.05}$ & $0.6_{-0.5}^{+2.7}$ \\
21477 & 10:00:22.15 & 2:23:56.10 & & 1.4 & 0.7 & & 2.474 & $10.57_{-0.02}^{+0.03}$ & $0.04_{-0.04}^{+0.18}$
\enddata

\tablecomments{Main properties of the Blue Jay quiescent sample. $U-V$ and $V-J$ rest-frame colors are computed using the EA$z$Y code (see \S~\ref{sec:uvj}); redshift, stellar mass and star-formation rates are estimeted by the stellar continuum Prospector fits (\S~\ref{sec:slit_loss_flux_calibration}).}

\end{deluxetable*}

\subsection{Alternative sample selection} \label{sec:uvj}
\begin{figure}
    \centering
    \includegraphics[width=\linewidth]{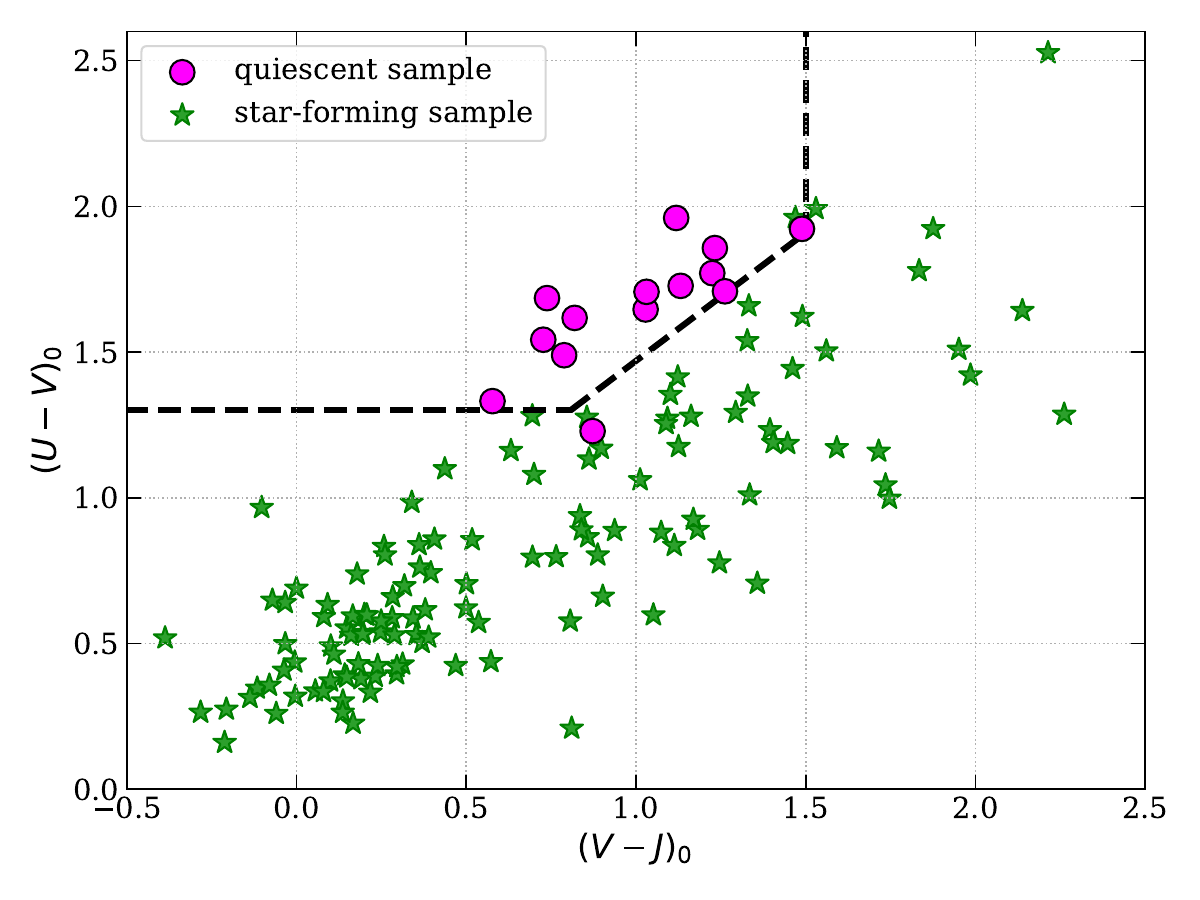}
    \caption{The Blue Jay sample on the $UVJ$ diagram. Dashed black lines mark the \cite{Muzzin13} selection of quiescent galaxies.}
    \label{fig:uvj_og}
\end{figure}

We check our selection by plotting the sample onto the rest-frame $UVJ$ color-color diagram (Figure~\ref{fig:uvj_og}), which has been proven effective in selecting quiescent galaxies up to \(z=3.5\) \citep{Williams09,Whitaker11,Straatman16}. We calculate the rest-frame colors using the public photometric redshift EA$z$Y code \citep{eazycode} and running it on the same photometry used for the Prospector fits (see \S~\ref{sec:slit_loss_flux_calibration}). The rest-frame $U-V$ and $V-J$ colors are reported in Table~\ref{prospector}. 

As expected, most of the selected galaxies are found inside the quiescent selection limit at $z>1$ \citep{Muzzin13}, traced by the black dashed line, with some galaxies found right on or just outside the boundary. A criterion based purely on the $UVJ$ colors would have selected a subset of our sample (see also discussion in \cite{Leja19} and \cite{Tacchella22}); the SFR-based selection can therefore be considered slightly broader. Given that the two types of selection result in similar samples, we conclude that our results could also be applied to a sample of color-selected quiescent galaxies for which no spectroscopy is available.

\begin{deluxetable}{cc}
%% This is the title of the table.
\tablecaption{Fitted emission lines \label{tab:lines}}

\tablehead{\colhead{Species} & \multicolumn{1}{c}{$\lambda_{rest-frame}$} \\ 
\colhead{} & \multicolumn{1}{c}{(\r A)}} 

%% All data must appear between the \startdata and \enddata commands
\startdata
$[\text{OII}]$ & 3727.10, 3729.86  \\
$[\text{NeIII}]$ & 3869.86 \\
 H$\beta$ & 4862.71 \\
 $[\text{OIII}]$ &4960.30, 5008.24 \\
 $[\text{NII}]$ & 6549.86, 6585.27 \\
 H$\alpha$ & 6564.60 \\
 $[\text{SII}]$ & 6718.29, 6732.67 \\
He I & 10833.31 \\
\enddata

%% Include any \tablenotetext{key}{text}, \tablerefs{ref list},
%% or \tablecomments{text} between the \enddata and 
%% \end{deluxetable} commands

\tablecomments{Rest-frame wavelength in vacuum.}

\end{deluxetable}

\section{Emission line fitting} \label{sec:fitting}

We employ the wide-range \texttt{Prospector} fits (\S ~\ref{sec:data}) to subtract the stellar contribution from the spectra and isolate the emission due to ionized gas. We then construct a model for the ionized gas spectrum, composed of a Gaussian profile for each of the emission lines listed in Table~\ref{tab:lines}.
Given the relatively low spectral resolution of the observations and the fact that we are observing gas-poor, quiescent galaxies, we adopt a single kinematic component with the same velocity dispersion and redshift for all the lines in each galaxy. As will be explored in \S~\ref{subsec:outflows}, these assumptions do not hold up for all galaxies in the sample. The flux ratios between [OIII]5007/[OIII]4959 and [NII]6581/[NII]6548 are fixed at the theoretical value of 3:1, following \cite{osterbrockferland}. Although our spectral resolution is not sufficiently high to be able to resolve the [OII]$\lambda\lambda$3727,3739 doublet, we fit the lines separately and impose the $1.5<$[OII]3729/[OII]3727$<2.5$ theoretical limit on their flux ratio \citep{OsterbrockIloveU}. 
The velocity dispersion parameter is fitted taking into account the contribution to the overall broadening of the lines given by the wavelength-dependent NIRSpec nominal spectral resolution.

The fitting is performed using the \texttt{EMCEE} Python library, an Affine Invariant Markov Chain Monte Carlo (MCMC) Ensemble sampler \citep{emcee}. We cut spectral regions of width 600 \r A in the observed frame around each emission line  and fit only the data within these regions. The free parameters of the model are the global redshift $z_\mathrm{gas}$ and velocity dispersion $\sigma_\mathrm{gas}$ (identical for all lines), and the flux of each line (apart from lines with a fixed flux ratio). Flux and velocity dispersion priors are constructed by employing flat logarithmic probability distributions, in order to represent our ignorance of the real probability distribution of these parameters. The velocity dispersion prior is set between 10 and 1000 km/s: given that we already account for the broadening due to NIRSpec spectral resolution, the lower and upper limits are physically-motivated to reproduce the typical range of observed velocity dispersions in line-emission distant galaxies. For the redshift parameter we adopt a Gaussian prior, centered around the \texttt{Prospector}-estimated value, with dispersion informed by the \texttt{Prospector} posterior distribution for $z$. The resulting best-fit emission line parameters are determined as the median values from the \texttt{EMCEE} posterior distribution, with uncertainties calculated as the values at the 16th and 84th percentile ranges, and are reported in Appendix \ref{appA}.

As already discussed in Section~\ref{sec:slit_loss_flux_calibration}, a correction is applied to the spectra in order to align the spectroscopic flux with the measured photometry. Assuming that the ionized gas emission is originating mainly in HII regions scattered throughout the galaxy, the slit-loss calibrated spectrum yields accurate flux estimations. However, it is possible that emission in these quiescent galaxies comes from limited regions of ionized gas that are not evenly distributed throughout the galaxy, e.g. only the nuclear regions: in this case, slit-loss flux corrections may not be accurate and fluxes may be overestimated.

\begin{figure*}[p]
    \centering
    \includegraphics[width=\textwidth]{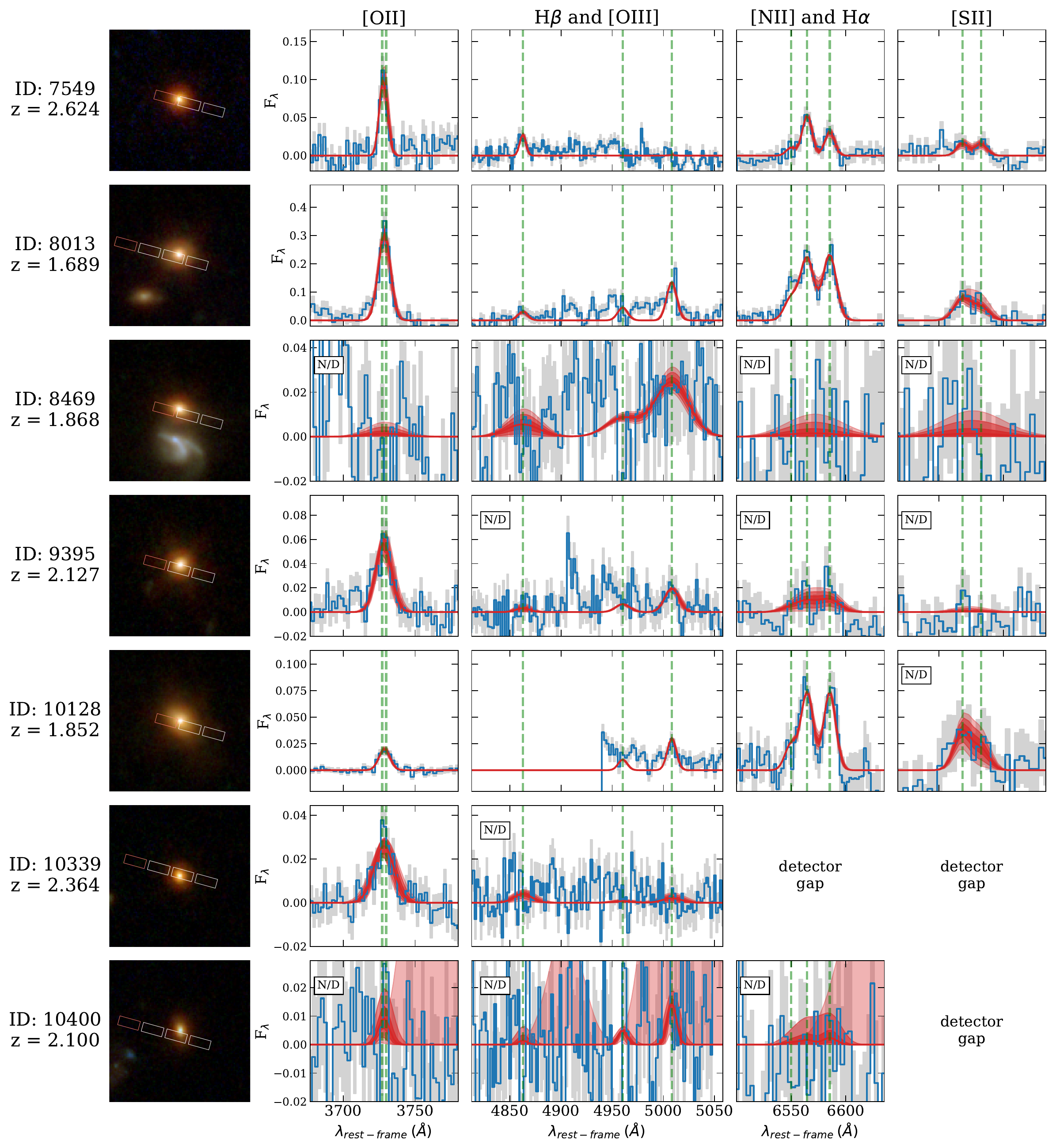}
    
    \label{fig:all_lines1}
\end{figure*}

\begin{figure*}[p]
    \centering
    \includegraphics[width=\textwidth]{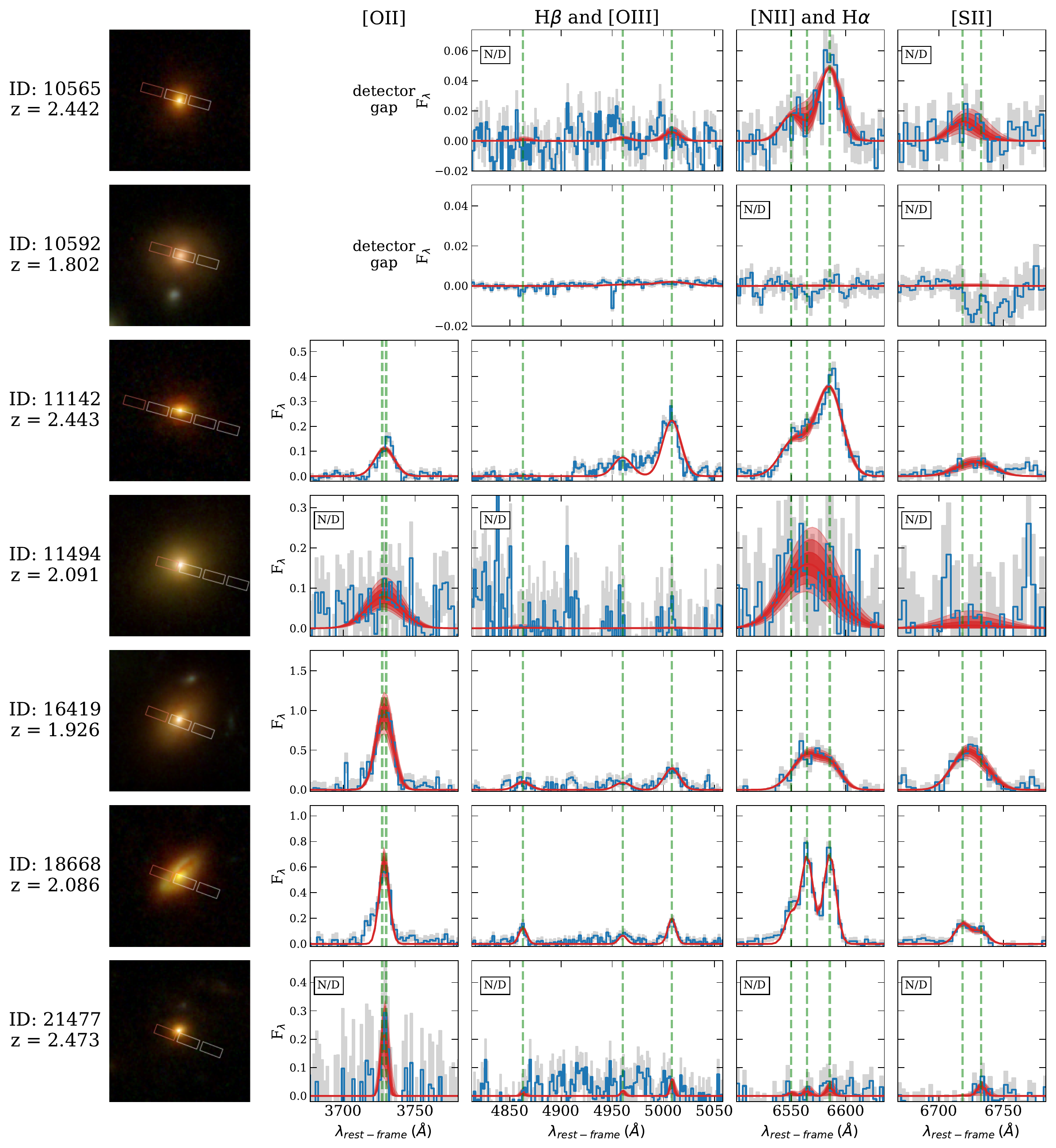}
    \caption{Cutouts and observed spectra for the sample of quiescent galaxies in the Blue Jay survey. We show $3" \times 3"$ RGB cutouts from NIRCam imaging data overlayed with the footprint of the MSA shutters, plotted in red and white for the A and B nod positions respectively. In the remaining panels, the observed spectrum is in blue, with the 1$\sigma$ uncertainty marked by the shaded grey region. Flux is expressed in $10^{-18}$ erg s$^{-1}$ cm$^{-2}$ \AA$^{-1}$ and given in observed-frame wavelength units, while the x-axis is converted in rest-frame wavelength for ease of viewing. The red line and shaded regions show the median and the $\pm1,2,3\sigma$ confidence regions for the results of the emission line fit. Panels where emission lines are not detected are labeled with ``N/D''. The reported redshifts are derived from the emission line fits.}
    \label{fig:all_lines2}
\end{figure*}

\section{Ionized gas content of quiescent galaxies} \label{sec:results_1}

The deep JWST/NIRSpec data obtained by the Blue Jay survey reveals the widespread presence of ionized gas in the quiescent galaxy population. Figure~\ref{fig:all_lines2} illustrates the HST cutouts and NIRSpec data for each galaxy. The continuum-subtracted data are shown in blue, while the best-fit spectrum obtained from the MCMC fitting is traced by the red solid line, with the red shaded regions indicating the $1\sigma,2\sigma$ and $3\sigma$ confidence levels, from less to more transparent.
We detect at least one emission line with SNR$> 3$ -- formal limit of line detection for this work -- in 10 out of 14 galaxies (71.4\% of the sample); and for most galaxies we find that multiple emission lines are robustly detected. This suggests the presence of continued ionizing sources in the population of quiescent galaxies at $z \sim 2$. Formally, only two spectra (IDs 10400 and 21477) out of 14 show no detected emission lines. However, we notice that the spectra of galaxies COSMOS-10592 and COSMOS-8469 appear to suffer from the presence of systematics in the data due to data reduction issues, which affected the stellar continuum fit: thus, the formal line detections we have for these galaxies are not reliable, and we treat them as non-detections.
Furthermore, we find that galaxy COSMOS-11494, the most massive in the sample, shows few emission lines that are formally detected but at very low SNR. Low SNR lines and non-detections characterize the noisiest spectra of the sample: in many of these cases, the MSA slits appear to be not well-centered on their galaxies: this could be a factor contributing to the missing detections of emission lines in these systems. In fact, the mean flux correction applied to the sample's spectra due to slit loss is, generally, of a factor $\sim2$: for the low SNR galaxies, however, the corrected flux averages at around 4 times the observed flux. 

Lines detected with the highest SNR in the sample are low-ionization lines, such as the [OII] doublet at $\lambda 3727,2729$\AA\ and H$\alpha$, while higher ionization lines such as [OIII]$\lambda$5007 are more rarely detected. The [NeIII] line, a high-ionization line which is considered to be a tracer of AGN emission, is successfully detected in only two galaxies (COSMOS-8013 and COSMOS-11142) and is thus not included in Figure~\ref{fig:all_lines2}. The He I line, which is a high-ionization line usually not detected in quiescent galaxy's spectra even at low-z, is actually detected with SNR$>$3 in 5 galaxies of the sample and is further discussed in Section \ref{sec:heI}.

\subsection{Emission line ratios} \label{subsec:bpts}

\begin{figure}
    \centering
    \includegraphics[width=\linewidth]{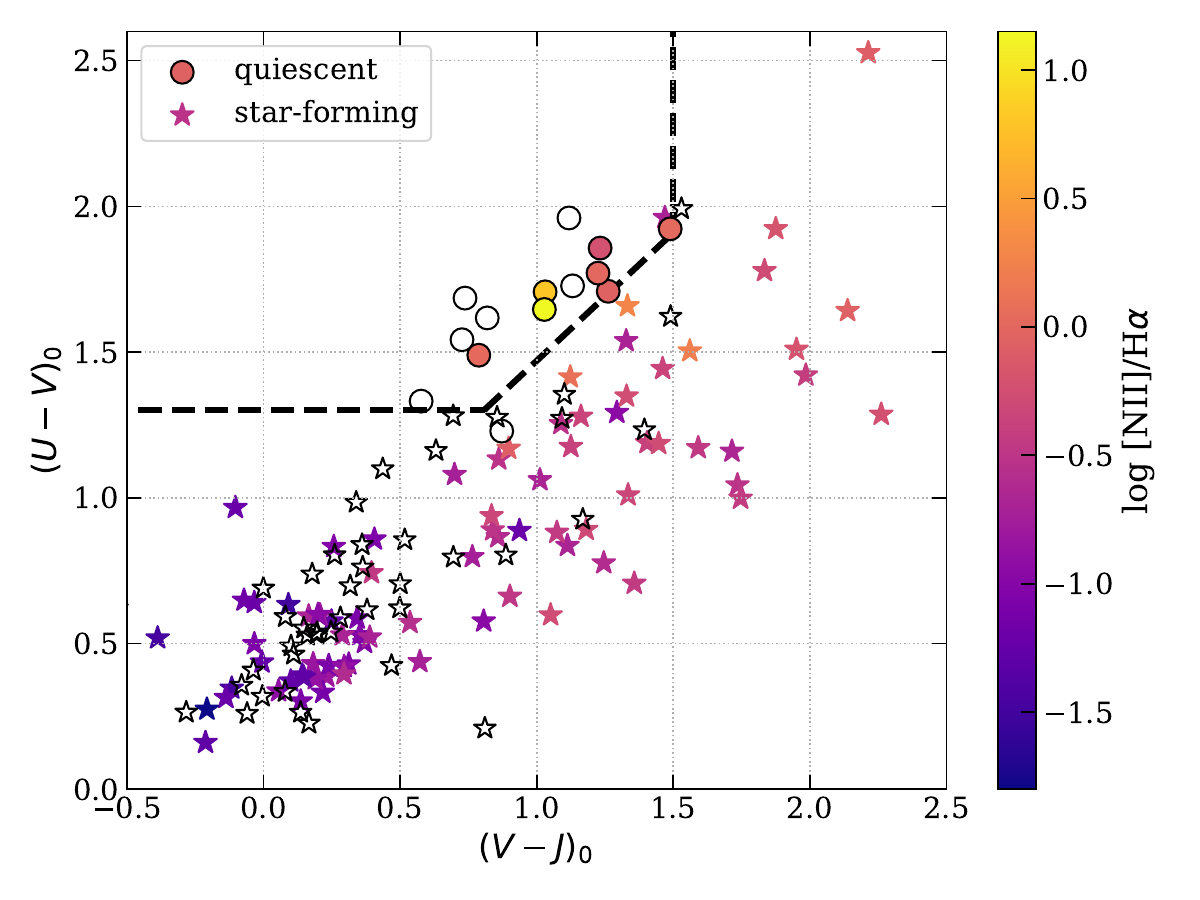}
    \caption{Blue Jay sample on the $UVJ$ diagram. The symbol shape indicates the SFR-based selection of quiescent galaxies, while the color coding shows the measured [NII]/H$\alpha$ line ratio. An empty marker means that [NII]$\lambda6583$ and H$\alpha$ are not detected with SNR$>3$.}
    \label{fig:test1}
\end{figure}

From Figure~\ref{fig:all_lines2} it is clear that, for most galaxies, the [NII]$\lambda6783$ line is similar to, or stronger than, the H$\alpha$ line. A high [NII]/H$\alpha$ line ratio is indicative of a hard ionizing photon field, typical of AGN or shock-induced photoionization \citep{emission_lines_bible}, even though it is also influenced by metallicity. In Figure~\ref{fig:test1}, we show the [NII]/H$\alpha$ ratio of the star-forming and quiescent galaxies in the Blue Jay sample, as a function of their position on the $UVJ$ diagram: most of the galaxies in the quiescent sample tend to have [NII]/H$\alpha$ ratio equal or higher than 1, while much lower values are typical of the star-forming sample. 
We also notice a trend in [NII]/H$\alpha$ ratio along the star-forming population, where redder galaxies tend to have higher values of the [NII]/H$\alpha$ ratio. This is likely due to the underlying correlation between color and mass: redder star-forming galaxies are more massive and thus contain more dust. Massive star-forming galaxies, in turn, have higher metallicity, resulting in a higher [NII]/H$\alpha$ ratio.
Moving towards the quiescent part of the diagram, however, we find generally higher [NII]/H$\alpha$ ratios compared to star-forming galaxies, suggesting a more complex origin behind the observed line ratio.

%discussion about the mass / NII /Halpha to be moved to a future paper??

\begin{figure}[t]
    \centering
        \includegraphics[width=\linewidth]{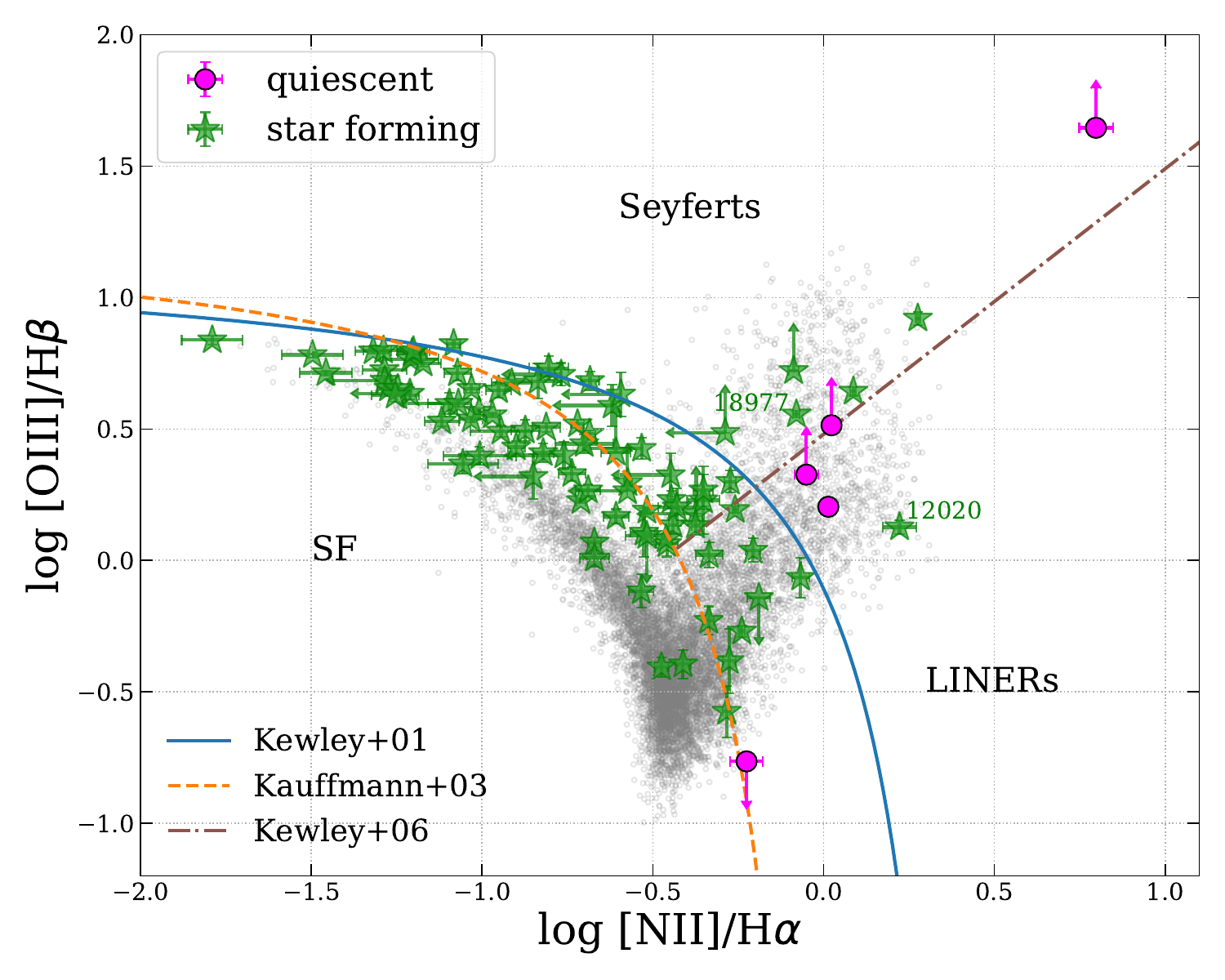}
        \caption{BPT diagram for the Blue Jay sample. Galaxies classified as quiescent from their SFR are shown with magenta circles. 
Star-forming galaxies are represented by green stars. The gray points are local galaxies from the SDSS \citep{Tremonti04}.}
        \label{fig:bpt}
\end{figure}

In order to explore the origin of the ionization, in Figure~\ref{fig:bpt} we show the Blue Jay sample on the [NII]/H$\alpha$ vs. [OIII]/H$\beta$ diagram, i.e., the so-called BPT diagram \citep{BPT_original}.
We consider galaxies where all four emission lines ([NII]$\lambda6583$, H$\alpha$, [OIII]$\lambda5007$, and H$\beta$) are detected with SNR $>$ 3; in order to include as many galaxies as possible from the quiescent sample, we also plot upper/lower-limit values at 3 sigma for systems with at least one detection at SNR $>$ 3 in each of the two line ratios.
Examples of galaxies added this way are COSMOS-8013 and COSMOS-16419. The distribution of the two samples on the diagram is compared with the observationally-derived star formation limit by \cite{kauffman03}, based on local SDSS galaxies, and the theoretical limit of extreme starburst galaxies by \cite{Kewley01}. The empirical classification of the AGN population into Seyfert and LINERs given by \cite{Kewley06} is shown as well. We find a clear segregation of the quiescent Blue Jay sample on the AGN-powered side of the diagnostic diagram, excluding one object (COSMOS-7549) which is instead found inside the extreme starburst limit, in the lower part of the diagram. This galaxy has non-detected [OIII] line, which translates into an upper limit on the [OIII]/H$\beta$ ratio which is much lower than what is found among the rest of the sample.

\begin{figure}[t]
        \centering
        \includegraphics[width=\linewidth]{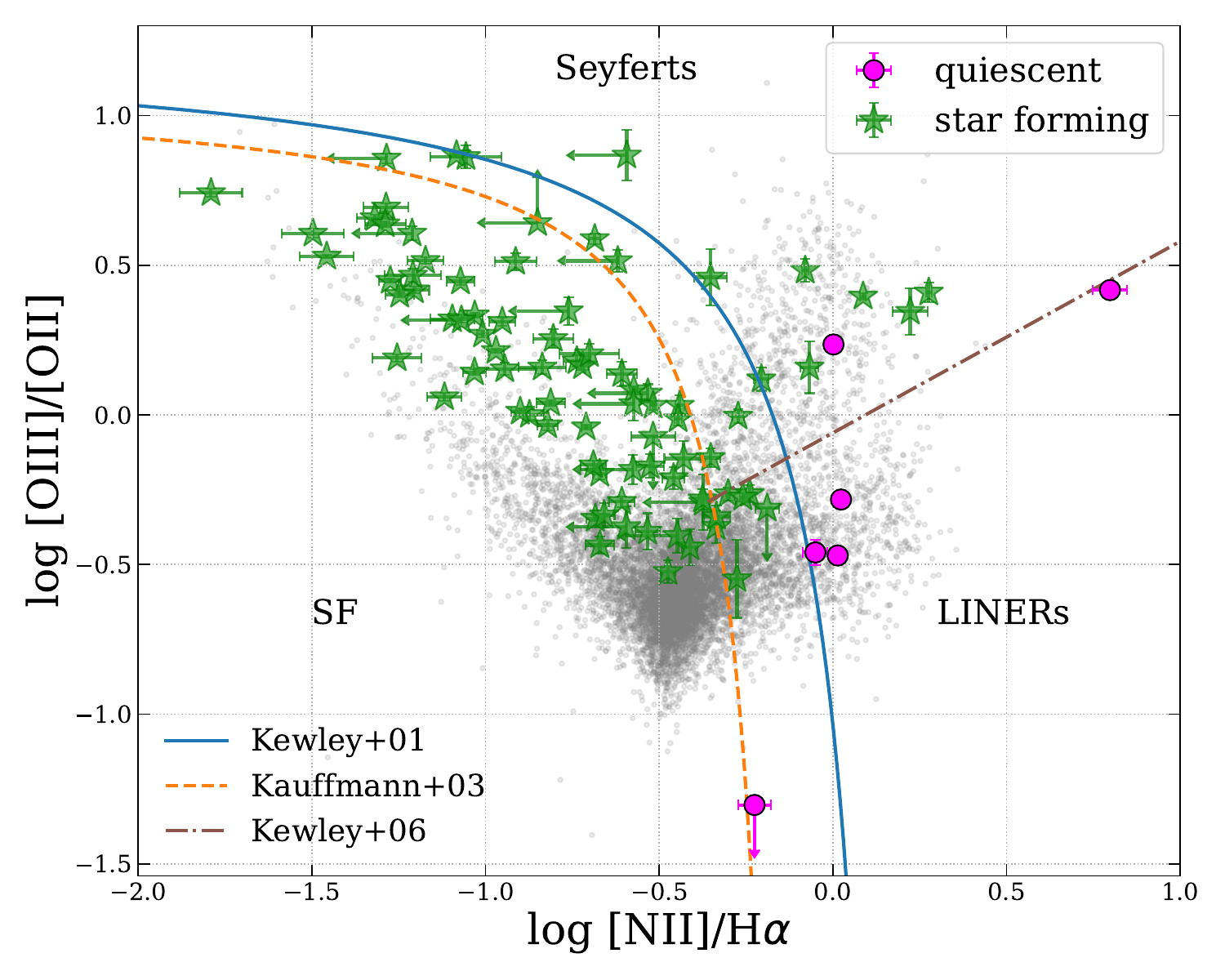}
        \caption{[OIII]/[OII] vs [NII]/H$\alpha$ diagram. Symbols as in Figure~\ref{fig:bpt}.}
        \label{fig:bpto2}
\end{figure}

\begin{figure}[t]
        \centering
        \includegraphics[width=\linewidth]{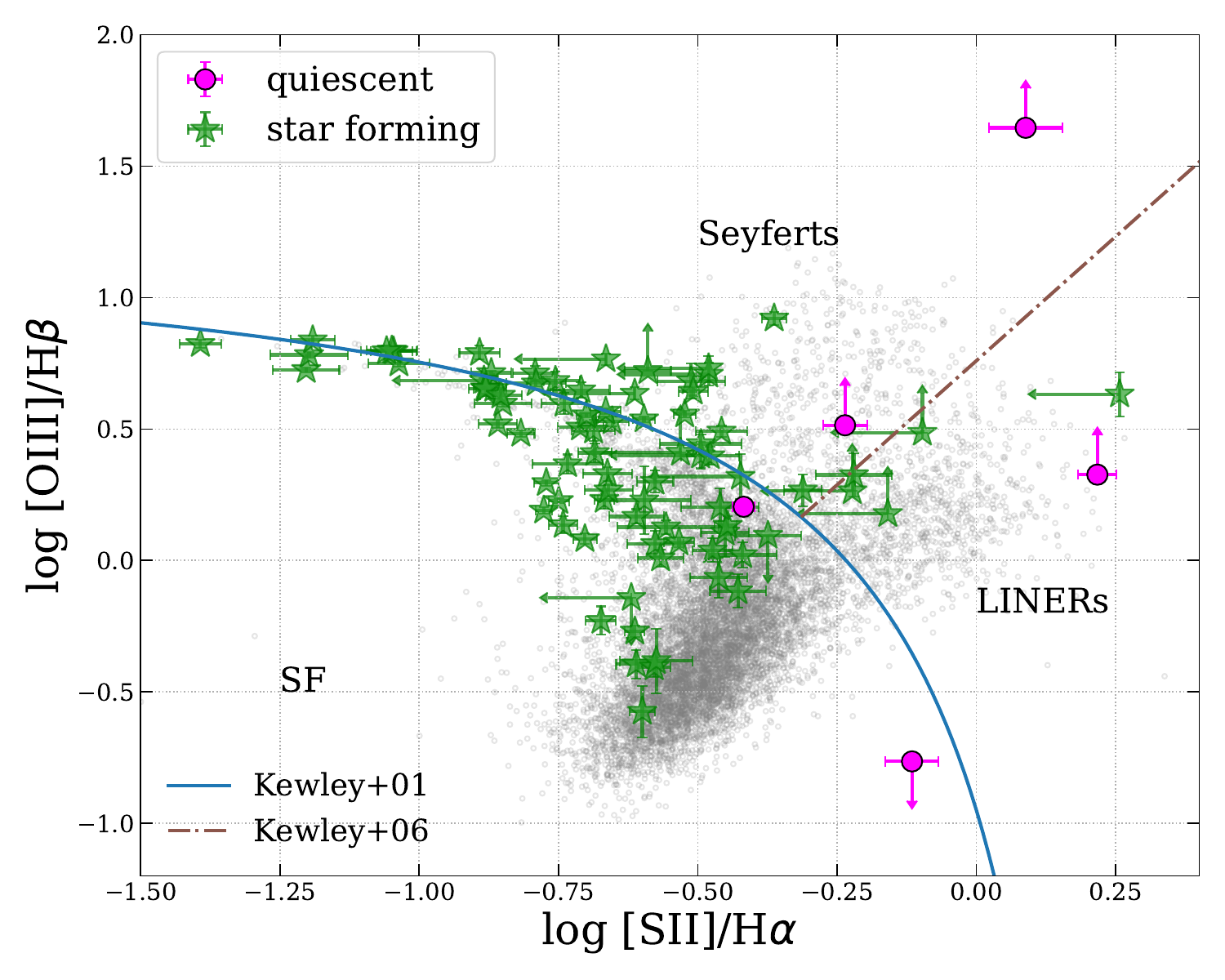}
        \caption{[OIII]/H$\beta$ vs [SII]/H$\alpha$ diagram. Symbols as in Figure~\ref{fig:bpt}. }
        \label{fig:bpts2}
\end{figure}

The star-forming galaxies in the Blue Jay sample form a well defined sequence along the so-called star forming galaxy abundance sequence \citep{Kewley06}. This sequence is slightly shifted towards higher [OIII]/H$\beta$ ratios compared to the distribution of local galaxies, which we take from the MPA-JHU DR7 release\footnote{Online repository: \hyperlink{}{https://wwwmpa.mpa-garching.mpg.de/SDSS/DR7/\#platelist}} of spectral measurements for the SDSS survey \citep{Tremonti04}. This is a well-known phenomenon observed at Cosmic Noon \citep{strom17,emission_lines_bible}, and is likely due to lower-metallicity and $\alpha$-enhanced stellar populations \citep{SANDERS20,Steidel16,Kewley13}. Despite this offset, the quiescent sample remains well-separated from the star-forming sequence, indicating that the primary ionizing source present in these galaxies is not star-formation, but rather the probable presence of an active galactic nucleus. This conclusion is further reinforced by the [SII]/H$\alpha$  \citep{VeilleuxOsterbrock87} and [OIII]/[OII] (not corrected for dust) \citep{Dopita2000} diagnostic diagrams, shown in  Figure~\ref{fig:bpts2} and Figure~\ref{fig:bpto2}: again, the quiescent sample is mostly found to the right of the maximum starburst limit, confirming that a surprising number of these galaxies may be hosting an AGN in their center. Interestingly, the star-forming galaxies (selected by their SFR) that are found on the AGN side of the diagnostic diagrams are a very small fraction of the total. Two of these AGN-hosting star-forming galaxies are COSMOS-18977 and COSMOS-12020, which are the only two spectroscopically confirmed Broad Line AGN in the Blue Jay sample (see Figure~\ref{fig:bpt}). 

    \vskip\baselineskip
    \hfill
\begin{figure}[t]
        \centering
        \includegraphics[width=\linewidth]{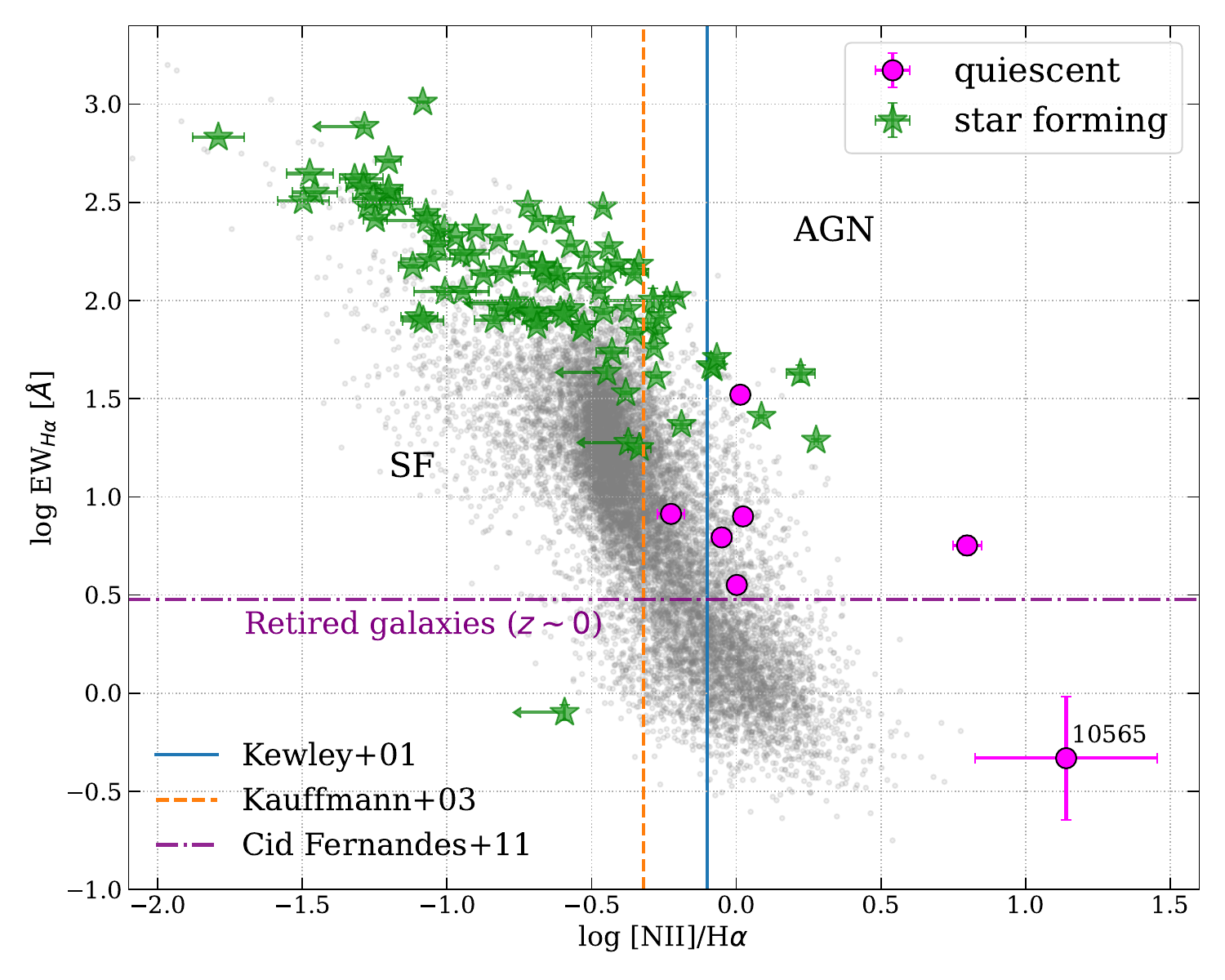}
        \caption{WHaN diagram. Symbols as in Figure~\ref{fig:bpt}.}
        \label{fig:whan}
\end{figure}

\subsection{H\(\alpha\) equivalent width}
\label{subsec:whan}
We also make use of the WHaN diagnostic diagram introduced by \cite{CideFernandes10,CidFernades11} (CF10,CF11). This line diagnostic diagram is similar to the BPT diagram, but with the [OIII]/H$\beta$ line ratio replaced by the H$\alpha$ equivalent width, which is easier to obtain in low-SNR spectra, thus allowing us to plot more quiescent galaxies with respect to the BPT. The results of the WHaN diagram, shown in Figure~\ref{fig:whan}, are consistent with those of the BPT diagram: we find most of the quiescent sample clustered on the AGN region of the plot. As before, the exception is COSMOS-7549, which is classified as a composite galaxy, as well as COSMOS-10565, which is the only one to be found in the "retired galaxies" region. 
When compared with the population of SDSS galaxies at $z = 0$, the distribution of the star-forming sample is systematically shifted towards higher H$\alpha$ EWs, as observed also by \cite{Belli17_kmos} at $0.7 < z < 2.7$: this shift can be attributed to the elevated specific SFR typical of Cosmic Noon galaxies \citep{MADAU2014}, which naturally results in a higher H$\alpha$ EW.

The WHaN diagram is very effective in distinguishing between low-power AGNs and weak Emission Lines Retired Galaxies (EL-RGs) (CF11). In retired galaxies the ionizing flux comes from hot evolved post-AGB stars \citep{stasinska06}, but due to the hard ionizing field these galaxies are often misclassified by the BPT diagram. To differentiate between true AGN and retired galaxies, CF11 propose a division on the WHaN diagram at EW(H$\alpha) = 3$~\AA. In our quiescent sample, only one galaxy (COSMOS-10565) falls below this line. However, we have to consider that the limit set at 3~\AA\ by CF11 has been calibrated on local galaxies and may not hold up at higher redshift. The quiescent galaxies observed by Blue Jay are substantially younger than those observed at $z \sim 0$ (many of them have stellar ages $\leq 1$~Gyr, see \citealt{MJ2024}), since the age of the Universe at $z \sim 2 $ is only 3 Gyr, and photoionization models show that the LIERs-like (Low-Ionization Emission Regions) line emissions due to post-AGB stars are actually weaker in younger stellar populations \citep{Byler19}. This means that the H$\alpha$ emission by retired galaxies at Cosmic Noon must be even fainter than the EW$ = 3$~\AA\ limit set by CF11. We thereby conclude that this type of ionization is not relevant for the vast majority of galaxies considered in this study.

\begin{figure}[h]
    \centering
    \includegraphics[width=\linewidth]{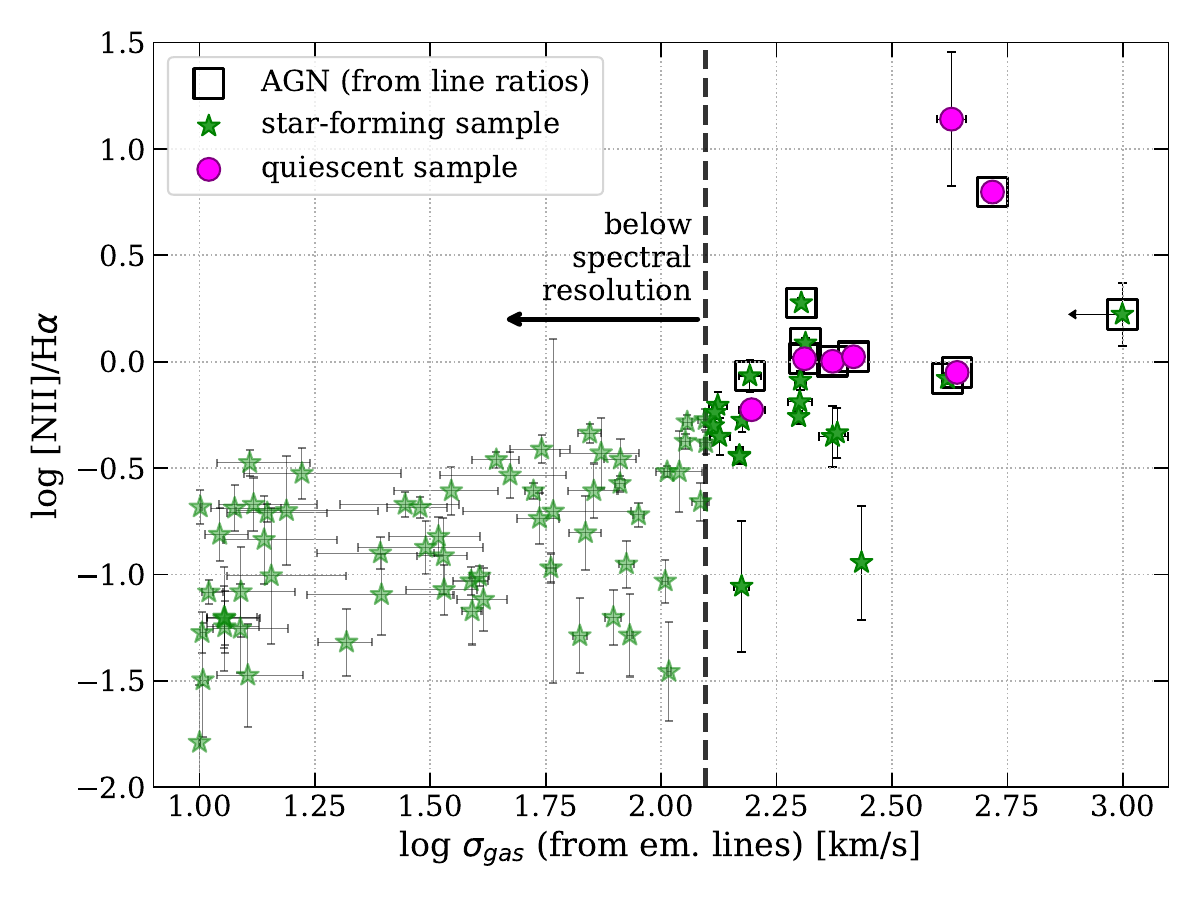}
    \caption{Comparison of the [NII]/H$\alpha$ ratio and the velocity dispersion of emission lines for the Blue Jay sample. Empty squares mark galaxies identified as AGN on the BPT diagram. For the broad-line AGN COSMOS-12020 we can only measure an upper limit on the velocity dispersion.}
    \label{fig:kinems}
\end{figure}

\subsection{Line kinematics \label{sec:kinematics}}
As a last piece of the puzzle, in Figure~\ref{fig:kinems} we show the relation between the [NII]/H$\alpha$ line ratio and $\sigma_\mathrm{gas}$, the velocity dispersion measured from the emission lines, for both the quiescent and the star-forming population. 
We find a clear division of the two samples, mirroring the division found in the other diagnostic diagrams: the quiescent sample is characterized by a higher [NII]/H$\alpha$ ratio and a higher gas velocity dispersion with respect to the star-forming sample. In many cases the velocity dispersion is so large ($\sigma_\mathrm{gas} > 400$~km/s) that it is clearly not due to the galaxy gravitational potential, but it must trace other processes such as outflows or shocks \citep{Rich2011,Rich2014}, which must be associated to AGN activity (as discussed below).
As a note, when fitting the velocity dispersions we adopted the nominal spectral resolution provided by JDox: however, we stress that the actual spectral resolution may be higher than this for compact sources (see \citealt{degraaf24}), thus the velocity dispersion values we provide may be underestimated. This issue mainly affects galaxies near or below the nominal spectral resolution, which is marked by the dashed black line in Figure~\ref{fig:kinems}, and is not  relevant for galaxies with high measured velocity dispersion. 
We also highlight with blue square frames the galaxies that are AGN hosts according to the diagnostic diagrams: notably, star-forming AGN hosts are found mixed in with the quiescent sample. Our analysis of the kinematics further confirms the AGN origin of the observed emission in quiescent galaxies.

\subsection{AGN activity in the sample \label{subsec:agns}}

The classification of galaxies in the sample based on their rest-frame optical lines is summed up in Table~\ref{tab:sfr}. We are able to classify 7 out of 14 galaxies (50\%), while the rest of them are either missing the required emission lines altogether or they have too low SNR for the diagrams to be reliable. Only one galaxy, COSMOS-7549, is classified as part of the composite population: this system completely lacks any [OIII] emission and may be a recently quenched non-active galaxy, whose residual ionizing flux from A-type stars is enough to produce low-ionization emission lines, such as [OII] and H$\alpha$, but no young stellar population is present to doubly ionize oxygen. This is similar to the sample of [OIII]-deficient galaxies analyzed by \cite{QuaiQuenching} at $z<0.21$; we further explore the star formation history of this system in Section \ref{sfr}.
Another galaxy, COSMOS-10565, is the only one classified as a retired galaxy by the WHaN diagram and could be a genuine inactive quiescent galaxy. However, its extreme [NII]/H$\alpha$ ratio and the presence of neutral gas outflows, as reported by \cite{Rebecca2024}, suggest a more complex reality. This specific system is further discussed in Section~\ref{subsec:outflows}.

Most importantly, all the other galaxies reveal signs of hosting an AGN in their center. AGN represent 71\% of the quiescent galaxy population for which we have reliable classification from line ratios, corresponding to a global 38\% of all galaxies in the quiescent sample. The incidence of AGN in our sample is comparable to the one found by \cite{forsterschreiber19} in gas-rich, massive star-forming galaxies at Cosmic Noon. However, this high incidence is not necessarily expected for gas-poor quiescent galaxies, such as the ones in our sample. These sources are not classified as AGN by any other catalog. None of these galaxies are detected in the X-rays by either the Chandra-COSMOS Legacy Survey Point Source Catalog \citep{chandracosmos} or the COSMOS XMM Point-like Source Catalog \citep{XMMcosmos}. None are detected in the radio as well, both by the VLA-COSMOS 3 GHz Large Project and the COSMOS VLA Deep surveys \citep{3ghzsurvey,VLAcosmos}. The AGN properties of the sample are discussed further in Section~\ref{subsec:agns_properties}.

One caveat is that the BPT and other rest-frame optical diagnostic diagrams cannot discriminate between shocks and AGN photoionization: \cite{dopita95_shocks} and \cite{Ho16} showed that diffuse radiation produced by fast shocks ($v_{gas} \sim 150-500$ km/s) from jets or star-formation feedback can reproduce optical emission line ratios of both LINERs and Seyfert galaxies. We are not able to rule out the possibility that shocks, rather than AGN photoionization, may play a dominant role in the excitation budget of the quiescent sample.  %, since we do measure gas velocities in the range of fast shocks ($180 \lesssim v_{gas} \lesssim 600$ km/s, where $v_{gas} = 0.5$FWHM(emission lines), see Appendix () for $\sigma_{gas}$ measurements). 
However, even if fast shocks were present in these quiescent galaxies, some of these quiescent galaxies also show outflows (see Section~\ref{subsec:outflows}) which cannot be star-formation-driven given the low SFR measured in these systems (see Section~\ref{sfr}). Moreover, from NIRCam images no AGN-hosting galaxy seems to be undergoing a merger or to have a nearby companion; we can thus exclude that the shocks are a consequence of galaxy interactions. The only possible source of shocks for the bulk of BPT-classified AGN, then, is through mechanical feedback from the AGN itself.  

\begin{deluxetable*}{cccccccc}
\tablewidth{2pt}
\tablecaption{Quiescent sample classification and SFR estimates \label{tab:sfr}}

\tablehead{\colhead{COSMOS ID} & \multicolumn{4}{c}{Classification from diagnostic diagram} & \colhead{Av} & \multicolumn{2}{c}{SFR (M$_{\odot}$/yr)} \\
\cline{2-5} \cline{7-8} 
\colhead{} & \colhead{BPT} & \colhead{WHaN} &
\colhead{[OIII]/[OII]} & \colhead{[SII]/H$\alpha$} & \colhead{}  & \colhead{from Prospector}  & \colhead{from H$\alpha$} } 

%% All data must appear between the \startdata and \enddata commands
\startdata
7549 & SF/Comp. & SF/Comp. & Comp. & SF & $-0.05_{-0.08}^{+0.07}$\tablenotemark{*} & $1.6_{-0.4}^{+0.6}$ & $0.145_{-0.003}^{+0.002}$ \\
8013 & Seyfert & AGN & LINER & Seyfert & $1.4_{-0.3}^{+0.5}$ & $0.07_{-0.07}^{+0.30}$ & $<2.0_{-0.7}^{+1.8}$ \\
8469 & $-$ & $-$ & $-$ & $-$ & $0.9_{-0.3}^{+0.4}$ & $0.02_{-0.01}^{+0.08}$ & $<0.06$\tablenotemark{\S}  \\
9395 & $-$ & $-$ & $-$ & $-$ &  $0.8_{-0.2}^{+0.3}$ & $0.1_{-0.1}^{+0.1}$ & $<0.14$\tablenotemark{\S}  \\
10128 & $-$ & AGN & Seyfert & $-$ &  $1.4_{-0.3}^{+0.4}$ & $2.6_{-0.4}^{+0.7}$ & $<0.6_{-0.2}^{+0.3}$  \\
10339 & $-$ & $-$ & $-$ & $-$ & $1.5_{-0.3}^{+0.4}$ & $0.03_{-0.03}^{+0.16}$ & $-$ \\
10400 & $-$ & $-$ & $-$ & $-$ &  $0.2_{-0.1}^{+0.3}$ & $0.003_{-0.003}^{+0.014}$ & $<0.02$\tablenotemark{\S}  \\
10565 & $-$ & EL-RG & $-$ & $-$ &  $1.3_{-0.3}^{+0.4}$ & $0.6_{-0.3}^{+0.4}$ & $<0.1_{-0.1}^{+0.1}$ \\
10592 & $-$ & $-$ & $-$ & $-$ &  $0.3_{-0.2}^{+0.2}$ & $0.007_{-0.007}^{+0.035}$ & $<0.001$\tablenotemark{\S} \\
11142 & Seyfert & AGN & LINER & Seyfert &  $2.4_{-0.4}^{+0.5}$ & $2_{-2}^{+4}$ & $<5.6_{-2.5}^{+4.9}$ \\
11494 & $-$ & $-$ & $-$ & $-$ &  $0.7_{-0.2}^{+0.2}$ & $0.2_{-0.2}^{+0.6}$ & $2.6_{-0.8}^{+1.1}$ \\
16419 &  LINER & AGN & LINER & LINER  & $0.4_{-0.1}^{+0.2}$ & $3.4_{-0.5}^{+0.6}$ & $<1.9_{-0.3}^{+0.5}$ \\
18668 &  LINER & AGN & LINER & SF & $3.04_{-0.13}^{+0.12}$\tablenotemark{*} & $0.6_{-0.5}^{+2.7}$ & $<45_{-6}^{+5}$ \\
21477 & $-$ & $-$ & $-$ & $-$ &  $1.3_{-0.5}^{+0.4}$ & $0.04_{-0.04}^{+0.18}$ & $<0.3$\tablenotemark{\S}
\enddata
\tablenotetext{*}{Dust attenuation derived from Balmer ratio}
\tablenotetext{\S}{H$\alpha$ line under detection limit}

%% General table comment marker
\tablecomments{Classification of the sample according to various diagnostic diagrams and SFR estimates from Prospector and H$\alpha$ line luminosity. Upper limits for SFR are given for galaxies with AGN contamination. Galaxy 10339 has no H$\alpha$ line due to a detector gap.}

\end{deluxetable*}

\section{Star formation rate} \label{sfr}
We derive again the SFR of each galaxy in the sample from the H$\alpha$ line luminosity, assuming the conversion provided by \cite{Kennicutt98}. However, the conversion of H$\alpha$ luminosity to SFR is affected by several issues.
First, we know from the emission line ratios that most of the H$\alpha$ emitters in the quiescent sample host an AGN, which is clearly the greatest contributor to the measured H$\alpha$ flux. In these cases, we consider the calculated SFRs as strict upper limits. 
Secondly, the measured line fluxes are affected by slit loss. This effect has been accounted for by the parametric correction estimated by Prospector and applied to the spectra before fitting the emission lines (see \S~\ref{sec:data}): as already mentioned in \S~\ref{sec:fitting}, however, the calibrated H$\alpha$ line luminosity is overestimated in the case of ionized gas emission whose origin is not evenly distributed throughout the galaxy. 
Finally, the measured line fluxes are affected by dust attenuation. Ideally, we would use the H$\beta$/H$\alpha$ Balmer decrement to estimate the nebular dust attenuation: in most cases, however, the SNR of the H$\beta$ line is too low. We therefore choose to estimate the attenuation $A_V$ using the results of the \texttt{Prospector} fits. The dust attenuation curve adopted by \texttt{Prospector} employs a two-components dust model \citep{CHARLOTFALL} consisting of a dust optical depth associated with diffuse ISM dust, plus an additional dust component associated with dense star-forming clouds. \texttt{Prospector} further models dust extinction using a power-law modifier $\delta$ to the \cite{CALZETTI2000} dust extinction law, following the empirical relation found in \cite{KRIEKCONROY}. Thus, we first employ the fitted two-component dust attenuation curve to derive the V-band extinction Av and then obtain the dust-corrected H$\alpha$ flux applying the modifier $\delta$ estimated by \texttt{Prospector} to the Calzetti law. 

\begin{figure}[h]
    \centering
    \includegraphics[width=\linewidth]{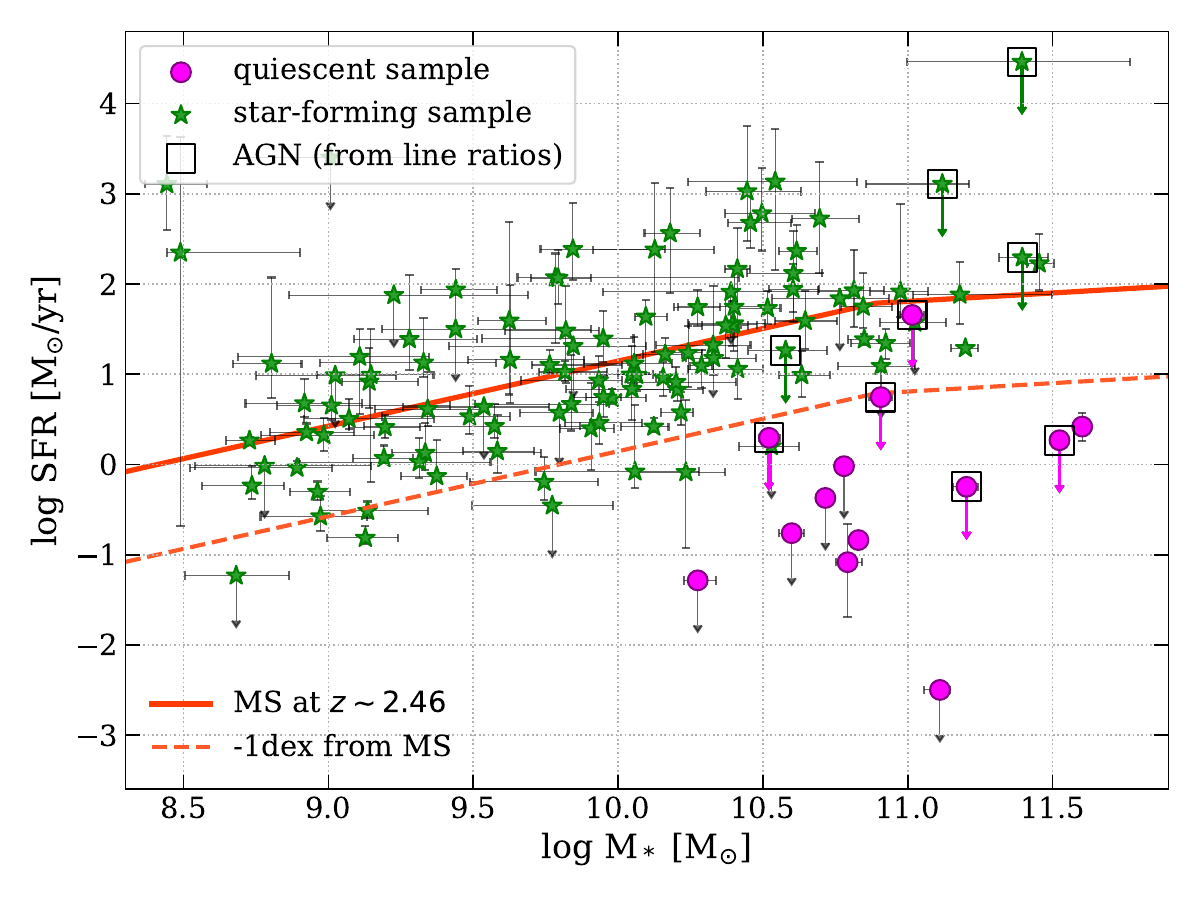}
    \caption{SFR from H$\alpha$ luminosity vs stellar mass for the Blue Jay sample. The orange solid line marks the star-formation main sequence at z = 2.46 from \cite{leja2022}; the orange dashed line shows the -1 dex limit. Empty squares mark galaxies classified as AGN hosts by the diagnostic diagrams. Black upper limits arrows are used for H$\alpha$ non-detections, while colored arrows indicate AGN contamination. }
    \label{fig:sfr_halpha}
\end{figure}

The SFR estimates based on the H$\alpha$ line luminosity for the quiescent sample are summarized in Table~\ref{tab:sfr}. In two cases (galaxies 7549 and 18668) we are able to recover the dust correction from the H$\alpha$/H$\beta$ ratio and then apply the Calzetti law directly, in all the other instances we use the Prospector dust model.  Figure~\ref{fig:sfr_halpha} depicts the position of the Blue Jay galaxies in the stellar mass vs SFR diagram, with the SFR of the star-forming sample computed in the same way. Galaxies hosting an AGN according to the line diagnostic diagrams are highlighted with black empty squares: for these cases, the computed SFR always represents an upper limit.  
We find results in agreement with the initial SFR vs stellar mass selection plot, which was based on Prospector measurements (Figure~\ref{fig:second_selection}). This confirms that the sample is indeed quiescent and lies well below the star formation main sequence at $z\sim2$.  Notably, one galaxy (COSMOS-18668) exhibit a SFR much higher than Prospector estimates, falling within $\pm 1$dex from the main sequence limit. However, given its status of AGN host, the measured H$\alpha$ emission predominantly originates from the active nucleus rather than star-forming regions, and thus its true SFR is lower than what measured from H$\alpha$. Interestingly, one of the two Broad Line AGN within the star-forming sample, COSMOS-18977 ($z=2.08,~\log(M_*/ M_{\odot})=10.70$), also borders the -1dex limit. For this galaxy, the stellar-continuum \texttt{Prospector} fit was performed only on the observed photometry \citep{MJ2024} due to the extreme broadening of permitted emission lines in the spectrum, thus its Propsector-estimated SFR may not be as accurate as for galaxies with a spectroscopic fit. In any case, we find a relatively low upper limit on its SFR from the H$\alpha$ line luminosity ($\text{SFR}\sim18\:M_{\odot}$/yr): given that the central AGN contributes significantly to the H$\alpha$ emission, and considering the presence of deep Balmer absorption lines in its spectrum, COSMOS-18977 could represent an additional massive quiescent galaxy hosting an AGN.

\subsection{Rejuvenated candidates}

\begin{figure}
    \centering
    \includegraphics[width=\linewidth]{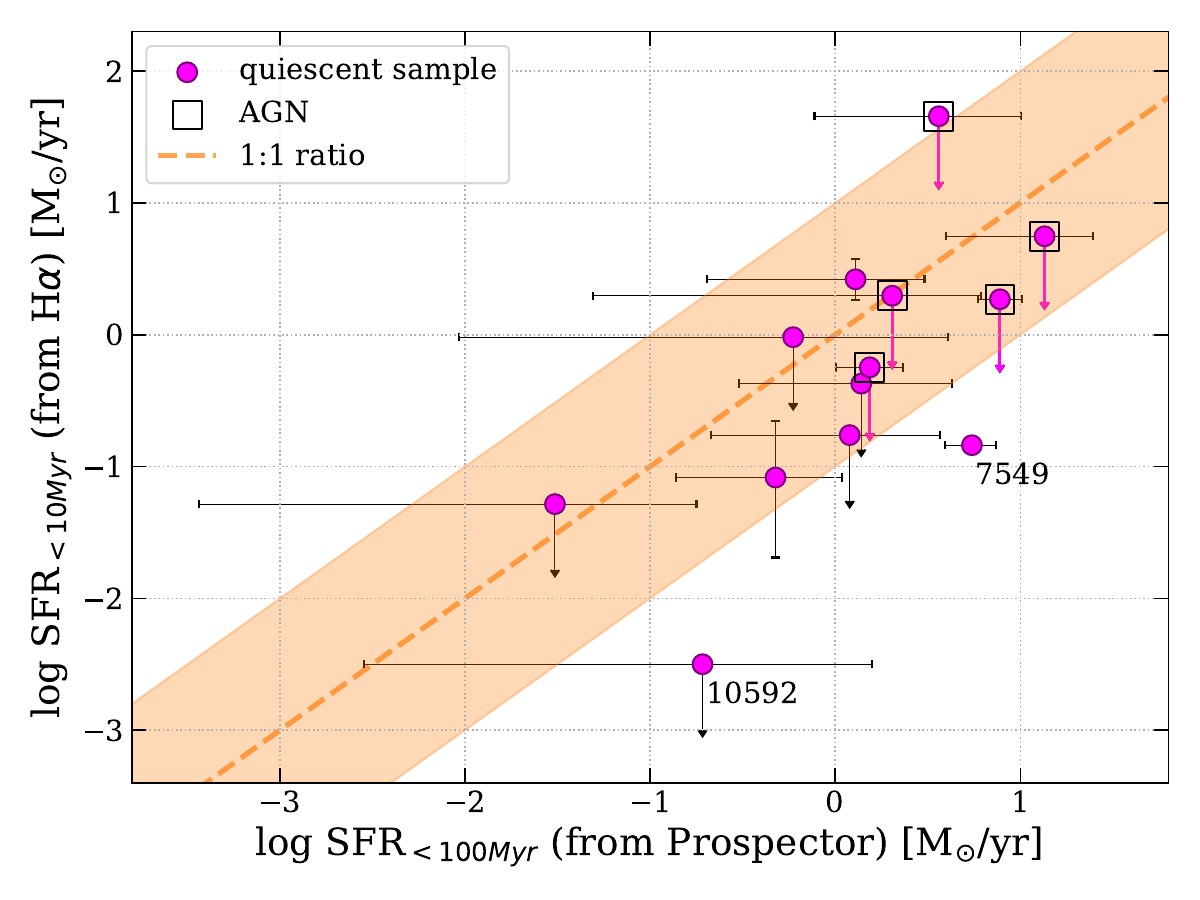}
    \caption{Comparison between the SFR computed on two different timescales: on a $\sim10$~Myr scale using H$\alpha$, and on a $\sim100$~Myr scale using \texttt{Prospector}. }
    \label{fig:sfr_comparison}
\end{figure}

The two SFR estimates of Table~\ref{tab:sfr} trace the star formation activity of galaxies on different timescales. The \texttt{Prospector} measurement corresponds to the SFR in the most recent bin in the SFH posterior distribution, which spans the last 30 Myr of the galaxy history. The stellar continuum, however, is not sensitive to changes in the stellar population on such short timescales. Moreover, the SFH fits employ a ``continuity'' Bayesian prior on the SFR between adjacent time bins, which disfavours abrupt changes in the SFH such as bursts, quenching and rejuvenation events on short timescales. On the other hand, the SFR estimated from the observed H$\alpha$ line luminosity traces very recent star-formation activity, since most of the H$\alpha$ flux in HII regions is due to short-lived massive stars, which photoionize the ISM around them on a time scale of $\sim 10$ Myr.

In Figure~\ref{fig:sfr_comparison} we compare the SFR measured with the two methods, in order to explore the possibility of \textit{rejuvenating} galaxies, i.e., galaxies that show recent star formation activity traced by H$\alpha$ but not detected by the \texttt{Prospector} fits. Because of the reasoning above, we choose to compute the \texttt{Prospector} SFR as the mean value over the last $\sim100$ Myr as obtained by the reconstructed SFH. We find only two obvious outliers in the SFR-SFR plot, with most galaxies found within 1~dex of a 1-1 relation between the two estimates (orange shaded region). One of the outliers is galaxy COSMOS-10592, whose spectrum suffers from systematic errors in the data reduction and for which we cannot trust the computed quantities. Apparently, no clear rejuvenated candidates are found in the sample, because no galaxy appears to have increased significantly its SFR in the last $\sim10$ Myr.

However, the second outlier, COSMOS-7549, may be of interest to understand variations of SFR on short timescales. This system is located at the highest redshift end of the sample ($z\sim2.6$) and is the only galaxy classified as part of the \text{composite} population by the diagnostic diagrams, meaning that its observed emission line ratios could be produced by a mix of recent SF and a contribution from a weak active nucleus or shocked gas emission. Thus, we would expect to observe a slightly higher SFR from H$\alpha$ in this system with respect to the others, but instead we find a very low rate of $0.145\:\: \text{M}_{\odot}$/yr --- one of the lowest of the sample. Furthermore, its H$\alpha$-estimated SFR is much lower than the \texttt{Prospector}-estimated one, suggesting that the galaxy underwent rapid quenching sometimes between $\sim100$ and $\sim10$ Myr ago. According to the SFH fits of \cite{MJ2024}, when using a ``bursty'' prior for the \texttt{Prospector} fits --- a modified prior which allows for more flexible SFR changes between time bins --- there is evidence for a recent rejuvenation event which briefly boosted the galaxy SFR, followed by rapid quenching. This scenario, then, could explain the observed emission lines: the rejuvenation event initially shifted the line ratios towards the star formation region in the BPT and WHaN diagrams, followed by a shift in the opposite direction as a consequence of the recent quenching. The final line ratios are thus found in the composite region. 
Interestingly, this may also explain the unusually low [OIII]/H$\beta$ and [OIII]/[OII] line ratios observed in COSMOS-7549: the high-ionization [OIII] line is expected to disappear after a few Myr since it requires very young and hot stars, while the lower ionization lines such as H$\beta$ and [OII] can be produced by B-type stars up to $\sim100$ Myr after the starburst event \citep{QuaiQuenching}. 
Therefore, in this scenario the small amount of ionized gas would come from an extremely recent and abrupt quenching of a minor rejuvenation event.

\section{Ionized gas outflows} \label{subsec:outflows}
The fitting model employed in Section \ref{sec:fitting}, made up of single-Gaussian profiles for each emission line, was not able to successfully fit the spectra of 3 galaxies in the quiescent sample: specifically, galaxies 8013, 11142 and 18668, all AGN hosts (Table~\ref{tab:sfr}). As visible in Figure~\ref{fig:all_lines2}, this simple model is inadequate for an accurate representation of the data, suggesting that more complex physical processes are involved in producing the observed emission. The detailed study of one of the galaxies, COSMOS-11142, has revealed the presence of a powerful outflow, which is responsible for deviation from the simple Gaussian profile in the ionized gas emission lines \citep{11142_article}. We thus employ a multiple-Gaussian model for these four galaxies, in order to investigate the presence of possible ionized gas outflows. 

\begin{splitdeluxetable}{c|cccc|cBc|cccc|cBc|cccc|c}
%% This is the title of the table.
\tablecaption{Results of double-Gaussian components fit \label{tab:outflows}}

\tablehead{ \multicolumn{6}{c}{COSMOS-8013} & \multicolumn{6}{c}{COSMOS-11142} & \multicolumn{6}{c}{COSMOS-18668}  \\
\multicolumn{1}{c|}{line} & \colhead{$\Delta v$} & \colhead{$\sigma_{nrw}$} & \colhead{$\sigma_{brd}$} & \colhead{$v_{out}^{ion}$} & \multicolumn{1}{|c}{$v_{out}^{neut}$} &
\multicolumn{1}{c|}{line}& \colhead{$\Delta v$} & \colhead{$\sigma_{nrw}$} & \colhead{$\sigma_{brd}$} & \colhead{$v_{out}^{ion}$} & \multicolumn{1}{|c}{$v_{out}^{neut}$} &
\multicolumn{1}{c|}{line} & \colhead{$\Delta v$} & \colhead{$\sigma_{nrw}$} & \colhead{$\sigma_{brd}$} & \colhead{$v_{out}^{ion}$} & \multicolumn{1}{|c}{$v_{out}^{neut}$} } 

%% All data must appear between the \startdata and \enddata commands
\startdata
$[\text{OII}]$ & $-407_{-279}^{+192}$ & $124_{-58}^{+44}$ & $700_{-151}^{+160}$ & $1107_{-32}^{+127}$ & & 
$[\text{OII}]$ & $-729_{-129}^{+74}$ & $277_{-22}^{+16}$ & $66_{-39}^{+72}$ & $796_{-2}^{+90}$ & & 
$[\text{OII}]$ & $-267_{-104}^{+94}$ & $126_{-58}^{+51}$ & $638_{-63}^{+78}$ & $905_{-15}^{+41}$ \\
$[\text{OIII}]$ & $-726_{-161}^{+172}$ & $146_{-74}^{+89}$ & $952_{-62}^{+35}$ & $1678_{-137}^{+100}$ & & 
$[\text{OIII}]$ & $-566_{-102}^{+97}$ & $288_{-18}^{+9}$ & $882_{-61}^{+62}$ & $1449_{-35}^{+42}$ & & 
$[\text{OIII}]$ & $-274_{-480}^{+334}$ & $186_{-52}^{+54}$ & $908_{-146}^{+68}$ & $1182_{-146}^{+335}$ \\
H$\alpha$, $[\text{NII}]$ & $-322_{-228}^{+129}$ & $194_{-26}^{+22}$ & $619_{-127}^{+134}$ & $946_{-5}^{+96}$ & & 
H$\alpha$, $[\text{NII}]$ & $97_{-38}^{+48}$ & $243_{-51}^{+34}$ & $501_{-42}^{+56}$ & $597_{-80}^{+104}$ & &
H$\alpha$, $[\text{NII}]$ & $-447_{-249}^{+221}$ & $153_{-14}^{+12}$ & $832_{-120}^{+107}$ & $1279_{-114}^{+130}$\\
\hline
\tablenotemark{*}NaI~D & $-679_{-133}^{+103}$ & & $64_{-46}^{+99}$ & & $839_{-188}^{+203}$ & \tablenotemark{*}NaI~D & $-212_{-29}^{+28}$ & & $54_{-30}^{+43}$ & & $323_{-64}^{+85}$ & \tablenotemark{*}NaI~D & $-143_{-33}^{+31}$ & & $228_{-50}^{+44}$ & & $600_{-112}^{+98}$\\
\enddata

\tablenotetext{*}{Neutral outflows measurements taken from \cite{Rebecca2024}.}

\tablecomments{All measurements are in km s$^{-1}$. }

\end{splitdeluxetable}
 
\subsection{Emission line broad components} 
Recent works on rest-frame optical emission of outflowing gas in active galaxies \citep{Brusa15,Ubler23} employ a fitting model with three Gaussian components for each line: one systemic, narrow component tracing the narrow line region (NLR) and/or star-formation; one very broad, systemic component for permitted transitions, tracing the broad line region (BLR); and finally a third component free to vary in both line width and centroid position, tracing ionized outflows. This model is physically motivated, but suffers from a degeneracy between the two types of broad component. In principle, the degeneracy can be broken by comparing the profiles of different emission lines, since the BLR component is present only in permitted transitions, while the outflow can be traced by both permitted and forbidden transitions. Among the emission lines detected in our sample, the only permitted transitions are H$\beta$, H$\alpha$, and He~I. However, H$\beta$ and He~I are generally too faint for an accurate modeling of the line profile; moreover, He I shows unique kinematics due to resonant absorption (explored in \S\ref{sec:heI}). Finally, H$\alpha$ is partially blended with the [NII] doublet, which makes it impossible to tell the difference between a broadening due to the BLR component of H$\alpha$ or a broadening due to an outflow component underlying all three lines. Therefore, in absence of evident BLR emission (such as for COSMOS-18977, as discussed in \S~\ref{sfr}), we cannot break this degeneracy for the galaxies in our sample.
Since in some cases we clearly observe broad components in forbidden lines, we conclude that ionized outflows must be present, and we adopt a simpler fitting model made up of only two Gaussian components, as done in \cite{forsterschreiber19, Zakamska16} and \cite{davies20}.
For each emission line, we fit a narrow component at the galaxy redshift with a velocity dispersion $\sigma \leq 300$ km/s and then we add a broad component with $\sigma\geq 200$ km/s tracing outflows/BLR emission, free to vary with respect to the galaxy systemic velocity. The broad component is considered only for lines with generally higher signal-to-noise ratios, namely H$\alpha$, [OII], [OIII] and [NII] doublets. Given the relatively low spectral resolution of our data and their blended profile, we tie together the kinematics of both the narrow and broad components for the [NII] doublet and H$\alpha$ line complex during the fit. The kinematics of the each line in a doublet are also tied together; and the same flux ratio limits employed in Section~\ref{sec:fitting} are applied.  

We perform the MCMC fits in a similar way as for the single-Gaussian model (see \S~\ref{sec:fitting}). The broad component velocity shift $\Delta v$ is initialized with a flat prior between -1000 and 1000 km/s; the limits on the velocity dispersion prior are $10-300$ km/s for the narrow component ($\sigma_{nrw}$) and $200-2000$ km/s for the broad component ($\sigma_{brd}$).
The MCMC walkers for the narrow component are initialised in small regions around the best-fit values given from the previous, single-Gaussian fit, while generic starting-point values are given for the broad component walkers.

\begin{figure*}[t]
    
    \centering
    \includegraphics[width=\textwidth]{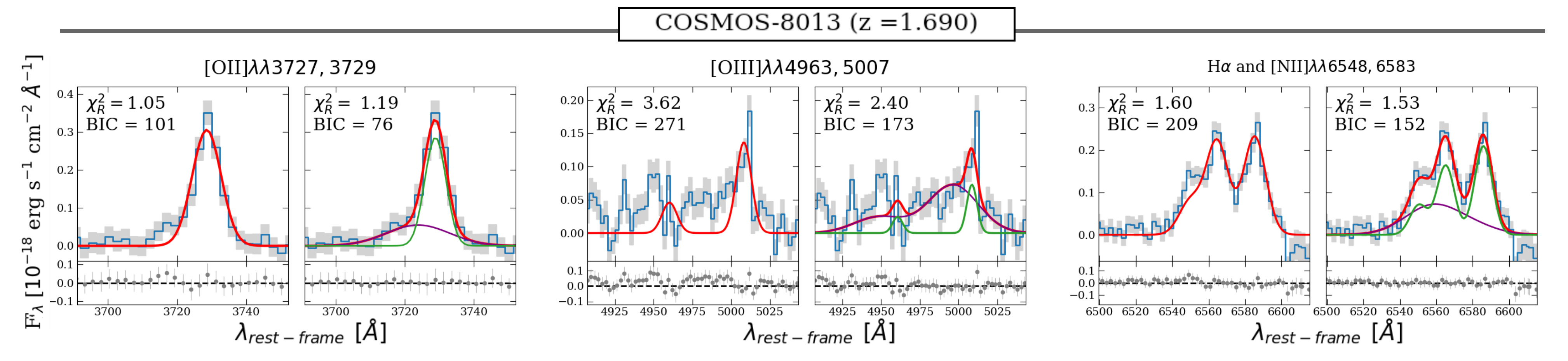}

    \centering
    \includegraphics[width=\textwidth]{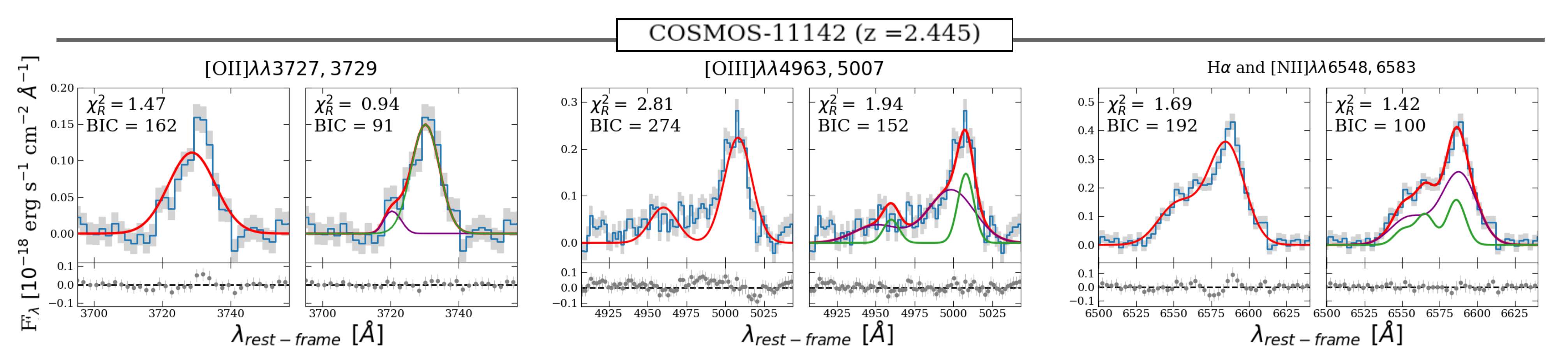}

    \centering
    \includegraphics[width=\textwidth]{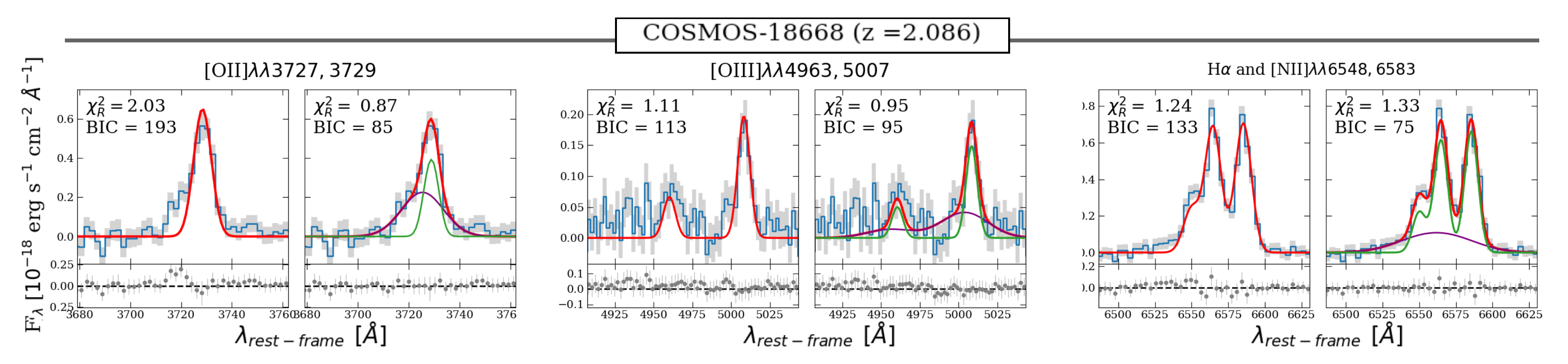}
    
\caption{Ionized outflows in three galaxies of the sample. For each galaxy and for each emission line group, we show a comparison between the single-Gaussian best fit (left-side panels) to the double-Gaussian best fit (right-side panels), complete with residuals (lower panels). In blue we plot the data, in grey shaded regions the $\pm1\sigma$ errors. Best-fit models are overlayed in red. For the double-Gaussian fits, we also plot the single components: in green the narrow one and in purple the broad.} 
\label{fig:outflows}
\end{figure*}

 The best-fitting parameters values for the double-Gaussian fits of are summed up in Table~\ref{tab:outflows}. In addition to these values, we also provide an estimation of the ionized outflows' velocity $v_{out}^{ion}$. In the case of a narrow cone-shaped geometry of the flow directed towards the observer, this would be computed as equal to the velocity shift $v_{out}^{ion} = |\Delta v|$; while in the case of a spherical geometry it is computed as $v_{out}^{ion} = |\Delta v| + 2\sigma_{brd}$. Since we do not know the geometry of the outflows, we choose to calculate the outflow velocity as an intermediate between the two, as done in \cite{11142_article}: $v_{out}^{ion} = |\Delta v| + \sigma_{brd}$.    
The results of the double-Gaussian model are compared with the single-Gaussian fit in Figure~\ref{fig:outflows}. For each of the three galaxies (rows), we show the single-Gaussian fit in the left panel and the double-Gaussian fit in the right panel for each emission line (columns) with residuals, along with the reduced-chi square values and Bayesian Information Criterion (BIC) values for each fit.  It is immediately visible that for the majority of emission lines, both residuals and goodness-of-fit metrics are greatly improved using the double-Gaussian model. Notably, each emission line in a single spectrum shows slightly different kinematics in its broad components, particularly in terms of velocity shift from the galaxy rest-frame. This changing behaviour is expected: in fact, differential (wavelength-dependent) dust absorption and the different ionization potentials of the lines can both greatly modify the shape of the outflowing (broad) emission. For example, broad emission in the bluer lines - such as the [OII] doublet - suffer more from dust extinction than in redder ones, rendering visible only the approaching end of the outflows and resulting in a more blueshifted and less broad line component. 
Finally, we note that the fitting of the broad component in the [OII] doublet of galaxy COSMOS-11142 was designed slightly differently from the rest of the fits: in this case, we widened the broad component's velocity dispersion prior range to 10 - 2000 km s$^{-1}$, in order to better fit the minor excess visible on the blue side of the line (Figure~\ref{fig:outflows}, left panels in the central row). This modification was motivated by inspecting the 2D spectra of this galaxy (see \citealt{11142_article}): we found that even though the bulk of emission from this line is aligned with the galaxy center, we can detect a very faint blueshifted, spatially misaligned component that we believe is at the origin of the slight blue excess visible in the 1D spectrum. 

 We interpret all of these blueshifted and/or broad emission components revealed by the double-Gaussian fits as outflows of ionized gas leaving the galaxies. The SFR we measure (see \S\ref{sfr}) is too low to explain the presence of star-formation driven outflows. We can also rule out a star-formation driven fossil outflow: in all galaxies we measure winds velocities of $\sim 800 - 1000$ km/s, substantially higher than what is observed in the most powerful star-formation driven outflows at $z\sim2$ \citep{davies19_outflows}. The high velocity, together with the lack of any tidal features visible in the near-IR imaging data, also rules out the possibility of tidally-ejected gas after a major merger.
The only reasonable origin for the observed outflows is AGN feedback, consistent with the position of these four galaxies on the line diagnostic diagrams.

\subsection{Multiphase outflows in quiescent AGN hosts}

Many galaxies in the Blue Jay sample host neutral outflows, as shown by an analysis of the Na~D absorption line conducted by \cite{Rebecca2024} on the same spectroscopic data. Significant blueshifted Na~D absorption, indicative of neutral outflows, is present in all three systems, as showed by the outflow parameters derived by \cite{Rebecca2024}, that we report in Table~\ref{tab:outflows} alongside the ionized gas measurements. 

The combination of ionized and neutral gas outflows in a subset of quiescent galaxies may provide a piece of evidence for negative AGN feedback on the star formation activity, since we are seeing gas leaving these already gas-poor, quiescent systems.  However, we have to be careful in tracing a link between observed multiphase gas outflows in this sample of quiescent galaxies to the quenching of star-formation in Cosmic Noon galaxies, given the difference in timescales between the two phenomena.

\begin{figure}
    \centering
    \includegraphics[width=0.9\linewidth]{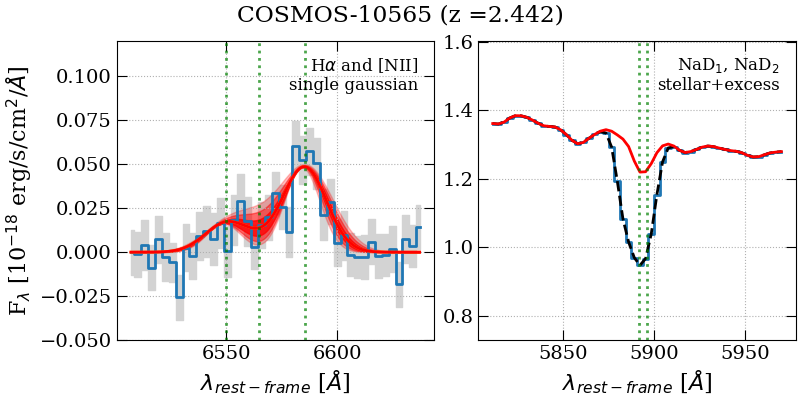}
    \caption{COSMOS-10565 is the only system in the sample to be classified as a retired galaxy by the WHaN diagram. Left: [NII] \& H$\alpha$ emission line fits; right: Na~D absorption line with best fit from \cite{Rebecca2024}. Despite the WHaN classification, the presence of a neutral outflow suggests that this galaxy hosts an AGN.}
    \label{fig:out10565}
\end{figure}

There are additional quiescent galaxies in which \cite{Rebecca2024} observe blueshifted Na~D absorption. Neutral outflows are detected even in galaxies that are not robustly classified as AGN according to their emission lines.
A particularly interesting case is that of galaxy COSMOS-10565 (Figure~\ref{fig:out10565}), which has weak emission lines, with the only detected ones being H$\alpha$ and [NII]. This system exhibits the highest [NII]/H$\alpha$ ratio of the whole sample and is the only galaxy classified as ``retired'' by the WHaN diagram (Figure~\ref{fig:whan}). The H$\alpha$ and [NII] line profiles are very blended and closely resemble the profiles observed in COSMOS-11142 (which is characterized by powerful outflows), but their SNR is too low for a meaningful fit with the double-Gaussian profile. If indeed an ionized outflow is present,  the outflowing gas is too faint to be seen robustly in the emission lines. 
The presence of a neutral outflow indicates that this galaxy is probably hosting a low-luminosity AGN, which is undetected due to the weak emission lines. Its low SFR of $\sim$ 0.1 M$_{\odot}$/yr (from H$\alpha$ line luminosity, Table~\ref{tab:sfr}) and the fact that it is isolated (Figure~\ref{fig:all_lines2}) rule out a major role for star formation feedback and galaxy interactions. 
This suggests that COSMOS-10565 is not a retired galaxy, despite its location in the WHaN diagram, and is consistent with the expectation that the local criterion from \cite{CidFernades11} (EW$_{H\alpha} < 3$~\AA) should be lowered at high redshift, as discussed in \S\ref{subsec:whan}.

\subsection{The He I line} \label{sec:heI}
\begin{figure*}
    \centering
    \includegraphics[width=\textwidth]{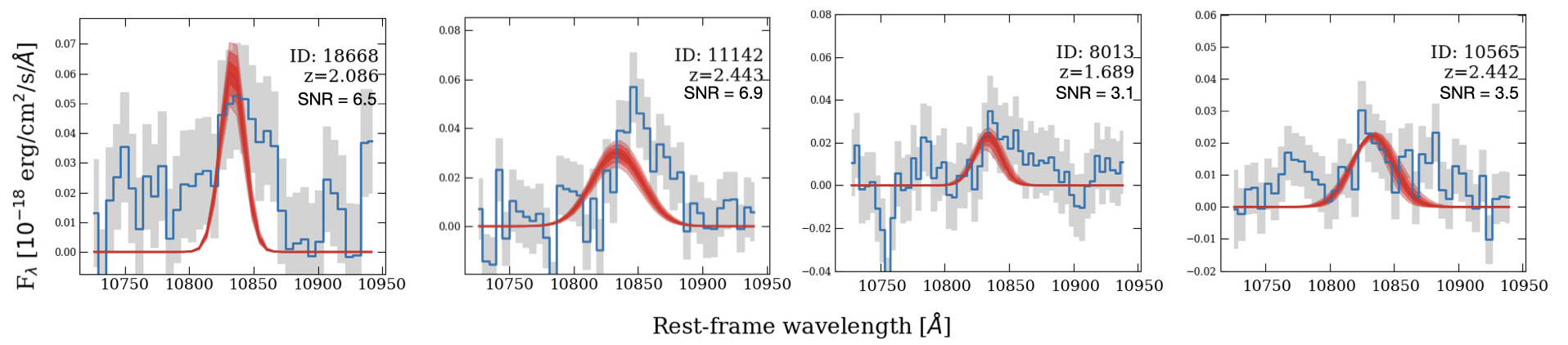}
    \caption{Detected He I $\lambda10831$ line in the quiescent sample. Data is in blue, with shaded gray regions indicating errors and single-profile Gaussian fits traced in red. Three galaxies (8013, 11142 and 18668) show evidence for excess flux on the red side, suggesting the presence of resonant scattering from the receding side of the outflow.}
    \label{fig:hei}
\end{figure*}

We report the detection of the high-ionization He I $\lambda10831$ \AA \hspace{1pt} line in 4 galaxies of the quiescent sample, shown in Figure~\ref{fig:hei} (we exclude a formal detection in COSMOS-10592 on account of the problems with the data, see \S~\ref{sec:results_1}). This emission line is detected exclusively in galaxies classified as AGN, all outflowing. 

In galaxy COSMOS-11142, which presents a wealth of ionized gas emission lines, the He~I line is the only one to be redshifted, as already found by \citet{11142_article}. Here we report two more cases -- COSMOS-8013 and COSMOS-18668 -- where we detected a red excess in the He~I emission line.
%As explained in \citet{11142_article}, this behaviour is likely a consequence of \textit{resonant scattering}, similar to that found in Ly$\alpha$ emission. This means that we are observing the redshifted (i.e. receding) part of the ionized outflows instead of the approaching one, due to the fact that the He I $\lambda$10831 transition is meta-stable and thus photons emitted in the outflows are re-absorbed unless they are redshifted. This constitutes an independent confirmation of the presence of ionized outflows in our sample of quiescent galaxies. 
As a caveat, we have to point out that in two of the cases (IDs 8013 and 10565) the He I line is only marginally above the nominal detection limit of SNR$>3$, as visible from Fig.\ref{fig:hei}: nevertheless, we choose to report these tentative detections anyway, on account of the novelty of observing this particular line in such a sample of high-z quiescent galaxies. The He I $\lambda10831$ line is not often detected in ISM studies, even at low-z, due to its relative faintness with respect to H I lines and its near-infrared central wavelength. However, He I has a higher ionization potential that H I (24.6 eV) and thus can be useful as an independent probe of a galaxy's multi-phase ISM. Since the launch of JWST, some studies have shown the usefulness of this line in detecting the presence of an AGN \citep{calabro23} or Wolf-Rayet stars  \citep{Morishita2024} in distant galaxies. 
 Whether it is redshifted or not, the He~I emission line in quiescent galaxies may represent a novel, promising indicator to investigate the presence of AGN activity at $z\sim2$.

\section{Summary and discussion} \label{sec:disc}
In this paper we analyzed the optical rest-frame emission lines of 14 quiescent galaxies observed as part of the Blue Jay survey, a medium-sized Cycle 1 JWST program that targeted 147 galaxies at redshift $1.7<z<3.5$ in the COSMOS field. The Blue Jay survey was executed with NIRSpec in MOS mode and obtained spectra between $1 - 5\mu$m, covering the $\sim3000-16000$ \AA\ rest-frame. The present work is the first statistical look at rest-frame optical ionized gas emission lines in a sample of massive quiescent galaxies at $z\sim2$ with deep spectra from JWST. All galaxies in the sample have stellar mass \(M_* > 10^{10}\: M_{\odot}\): in the Blue Jay survey, quiescent galaxies represent 23\% of all massive galaxies. The quenched fraction observed in the Blue Jay sample is consistent with other photometric samples at similar redshifts, such as \cite{Muzzin13} and \cite{sherman2020}: we refer to \cite{MJ2024}'s paper for a more in-depth breakdown of Blue Jay's quenched fraction and its trend with stellar mass.  
Quiescent galaxies were selected from the main sample by requiring their SFR, determined by spectro-photometric fitting, to be 1~dex below the main sequence. SFR upper limits determined by the H$\alpha$ line luminosity confirm that the sample is truly quiescent.  
Nonetheless, we detect emission lines from ionized gas in 10 out of 14 galaxies, representing 71.4\% of the sample. This shows that very deep spectroscopy is able to reveal the presence of ionized gas in the majority of high-redshift quiescent galaxies, confirming a trend established by increasingly deep ground-based observations such as the Heavy Metal survey \citep{heavymetalsurvey}, which detected H$\alpha$ emission in 55\% of their quiescent galaxy sample at $z>1.4$, and small samples of strongly lensed galaxies \citep[e.g.,][]{newman18}.

\subsection{AGN activity in quiescent galaxies at $z\sim2$ \label{subsec:agns_properties}}

We highlight the detection of multiple emission lines in addition to H$\alpha$ throughout the sample: the exquisite sensitivity of JWST/NIRSpec observations in the near-IR has enabled the detection of weak ionized gas emission lines that would have been otherwise missed by ground-based observations. This, in turn, made it possible to classify galaxies using line ratio diagnostic diagrams, which revealed rampant AGN activity in our sample of massive quiescent galaxies.
We find that 6 out of 14 galaxies show emission line ratios compatible with the presence of an active nucleus (43\% of the sample). We consider the possibility of contamination of the observed line ratios by fast shocks, but we find no possible sources of shocked gas other than AGN activity itself to be present in these galaxies. Interestingly, no X-ray or radio detection is found for any of the AGN in our sample. 
The main result of this work thus consists in the discovery of a population of never-before-detected AGN at $z\sim2$ in a sample of apparently inactive massive quiescent galaxies.

To investigate the AGN properties, we start by estimating the bolometric luminosity of each AGN from the observed [OIII] line \citep{Heckman04}.
Most of the AGN exhibit bolometric luminosities of the order of $\sim10^{44}$ erg/s; the largest value is found for COSMOS-11142 (L$_{BOL} = 8.2 \pm {0.2} \cdot 10^{44}$ erg/s).
We then derive the inferred X-ray luminosities in the 2-10 keV range from scaling relations \citep{Duras2020}.
We find all of the sample falls under the flux limits of both the Chandra-COSMOS Legacy Survey Point Source Catalog \citep{chandracosmos} and the COSMOS XMM Point-like Source Catalog \citep{XMMcosmos}. Furthermore, the observed [OIII] line luminosity may come also from shock excitation, thus both the bolometric luminosity and inferred X-ray luminosity have to be considered upper limits. We thus conclude that the lack of X-ray detections is expected, given the weak AGN luminosities.

\begin{figure}
    \centering
    \includegraphics[width=\linewidth]{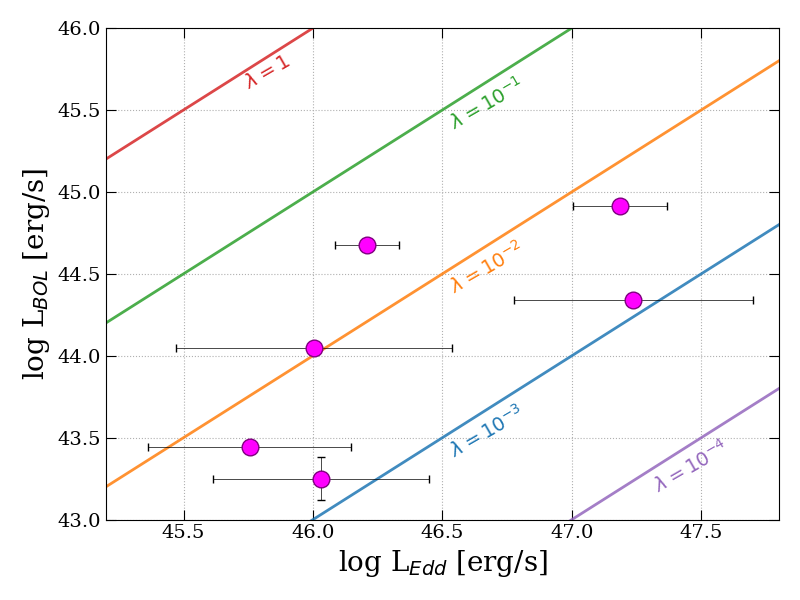}
    \caption{Bolometric luminosity (derived from the observed [OIII] emission line) versus Eddington luminosity (inferred from the stellar velocity dispersion) for the AGN hosts in the sample. Diagonal lines have constant values of Eddington ratio $\lambda$.}
    \label{fig:edd}
\end{figure}

\begin{figure*}[t]
    \centering
    \includegraphics[width=0.8\textwidth]{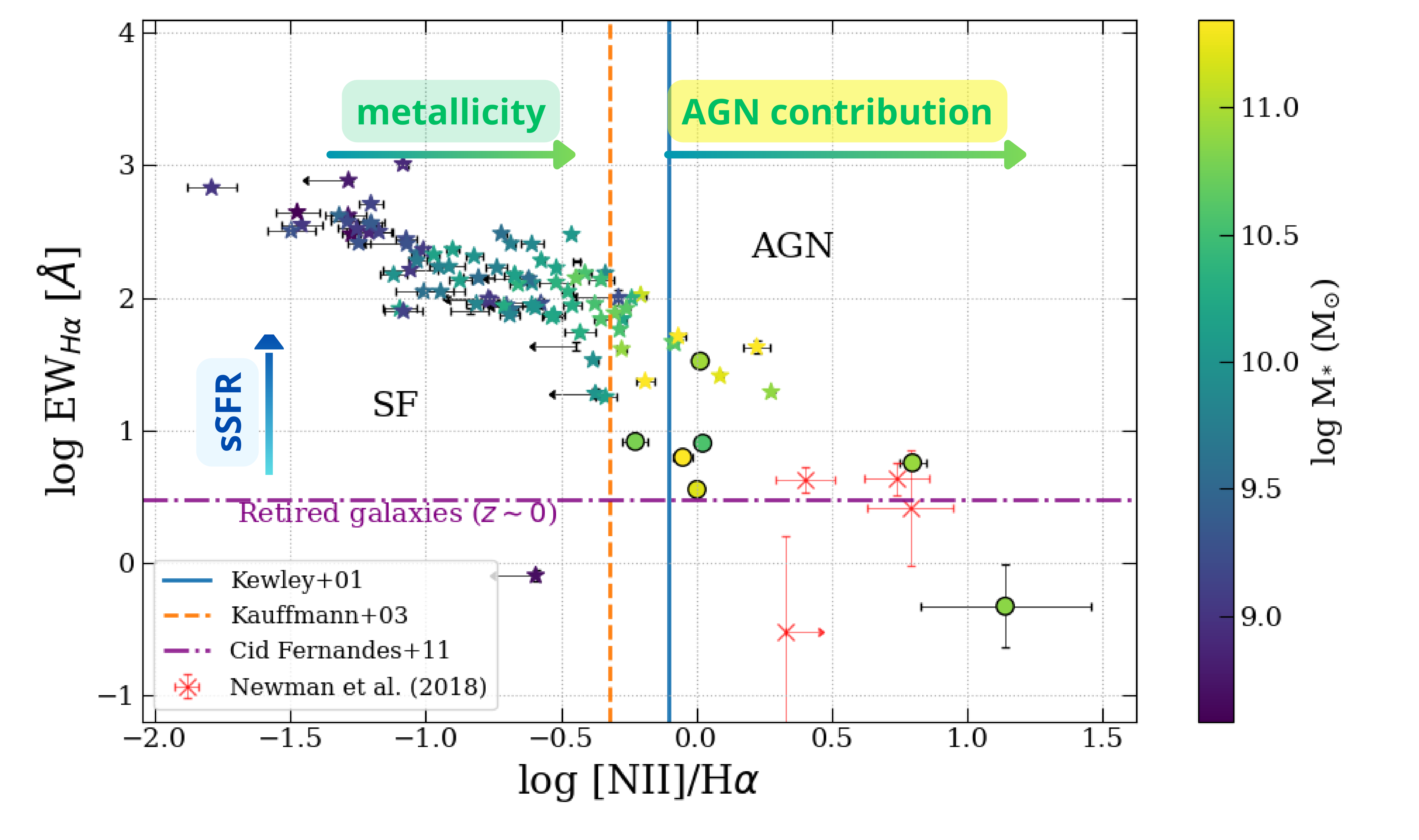}
        \caption{WHaN diagram, as the one shown in Figure~\ref{fig:whan}, but with symbols color-coded by stellar mass. Lensed quiescent galaxies from \cite{newman18} are also shown as red crosses. Qualitative trends with specific SFR, metallicity and contribution from AGN are indicated by the arrows. Galaxies in the Blue Jay sample appear to form a diagonal sequence: starting from the low-mass star forming population at the top left corner (stars), as galaxies grow in mass and turn into quiescent galaxies (circles), their specific SFR decreases and the contribution by LL-AGN to the line ratios increases.}
    \label{fig:whanarrows}
\end{figure*}

We also estimate the Eddington luminosity $L_{Edd}$ from the black hole mass, which we infer from the M$_{BH}-\sigma$ relation of \cite{McConnell2011}:
\begin{equation}
    \left ( \frac{M_{BH}}{10^8 \text{M}_{\odot}} \right )  = 1.95\left ( \frac{\sigma}{200\: \text{km s}^{-1}} \right )^{5.12} \;,
\end{equation}
adopting the stellar velocity dispersions measured from the Prospector fits. We show the bolometric luminosity versus the Eddington luminosity for the AGN hosts in our sample in Figure~\ref{fig:edd}, where the diagonal lines have constant Eddington ratio $L_{bol}/L_{Edd}$. Most of the AGN are probably low-luminosity AGN (LL-AGN), with Eddington ratios of $\sim 10^{ -3} - 10^{-2}$, typical of a central engine fueled by a Radiatively Inefficient Accretion Flow (RIAF) (see \citealt{Ho08LLAGN} for a schematic view of the central engine of LL-AGN). As already mentioned in \S~\ref{sec:kinematics}, the measured velocity dispersions may be underestimated, thus Eddington ratios may actually be lower than computed here. 
Even though the high incidence of AGN discovered in this sample of apparently inactive quiescent galaxies has not been seen before at $z\sim2$, this result is consistent with what is observed in the local Universe: \cite{Ho08LLAGN} analyzed optical emission lines in several surveys of nearby galactic nuclei, finding that 43\% of all galaxies show evidence of weak nuclear activity from an accreting SMBH and that the AGN fraction is even more remarkable for galaxies with an obvious bulge component, rising up to $\sim50\%-70\%$ for Hubble types E-Sb. Our measured AGN incidence of 43\%, then, is in line with what is observed at $z\sim0$, considering that galaxies with missing line detections could also be hosting LL-AGN out of the sensitivity reach for this survey. Within this framework, it is reasonable to think that many of the Blue Jay galaxies in the star-forming sample may also be hosting a low-luminosity active nucleus whose line emission is however outshined by the nebular emission from HII regions, a problem that does not affect the quiescent sample. To illustrate this idea more clearly, we can follow the relative contributions of star-formation and LL-AGN as a sequence across the WHaN diagram of Figure~\ref{fig:whanarrows}, where symbols have been colored by stellar mass. Gas metallicity also affects line ratios in galaxies of similar masses. In the top left corner (highest H$\alpha$ EWs and lowest [NII]/H$\alpha$ ratio) we find low-mass, low-metallicity SF galaxies, which probably do not host any AGN: at these low stellar masses, line ratios are determined by star-formation processes and the spread in [NII]/H$\alpha$ ratio values is likely due to different gas metallicity. Moving towards lower H$\alpha$ EWs, we find higher-mass galaxies: as the specific SFR decreases in the vertical direction, the contribution of a possible LL-AGN in the central regions raises the [NII]/H$\alpha$ ratio with respect to star-formation only emission, until the massive quiescent galaxies in the bottom right corner of the diagram are reached, for which the LL-AGN contribution is dominant. For high mass galaxies, metallicity should have a smaller impact on the line ratios, as shown by the relatively small difference in [NII]/H$\alpha$ ratio for galaxies of different masses at fixed H$\alpha$ EW. Additionally, we have plotted the position of the 4 lensed massive ($M_{*}>10^{11}\:M_{\odot}$) quiescent galaxies observed by \cite{newman18} at $z\sim2$: these galaxies have very similar properties to the Blue Jay quiescent galaxies and have also been observed to be AGN-hosts. Their position is very much in line with our quiescent sample and seems to also fill our "gap" in the [NII]/H$\alpha$ line ratio range. 
The observed sequence could thus be seen as an evolutionary track on the WHaN diagram, tracing the evolution of massive galaxies from star-forming to quiescent as driven by the effect of AGN feedback.

\subsection{Star-formation quenching by multiphase AGN outflows}

We find ionized outflows in a subset of 3 out of 14 quiescent galaxies, and measure outflow velocities of the order of $\sim1000$~km/s. All three galaxies are AGN hosts (according to the line diagnostic diagrams) and display evidence of neutral gas outflows as well \citep{Rebecca2024}. Additionally, one other galaxies categorized as AGN hosts exhibit solely neutral phase outflows (COSMOS-10565). The observed outflows can only be attributed to AGN activity.

One of the galaxies in our sample is COSMOS-11142, which was studied in detail in \citet{11142_article}. The mass outflow rate in this system is sufficiently large to fully explain the rapid shut-off of star formation, providing one of the first direct observational links between quenching and AGN ejective feedback. However, in this galaxy most of the gas that is being ejected is in the neutral phase, with the ionized phase playing a minor role. The other galaxies in our sample have ionized outflows that are comparable to, or smaller than, the one observed in COSMOS-11142. Thus, we do not expect that these quiescent galaxies have been quenched solely by the observed ionized outflows.
However, other galaxies in our sample do show evidence of neutral outflows, and are therefore similar in many respects to COSMOS-11142, even though none of them is close in terms of providing such a clear quenching picture. One possibility is that COSMOS-11142 was caught at the ideal time, just in the middle of the short-lived, powerful multiphase outflow episode that is responsible for rapidly quenching most of the star formation activity in the galaxy. In this scenario, the other galaxies in our sample are observed in later evolutionary stages: nonetheless, we find that the majority of them still host detectable AGN activity. Thus, our results might represent a new piece of the quenching puzzle.

Finally, it is possible that the observed LL-AGN activity plays a key role in maintaining quiescence in massive galaxies as they evolve to $z \sim 0$. In the local universe, a particular class of quiescent galaxies exhibiting LL-AGN activity and ionized gas winds, called ``red geysers'', has already been observed \citep{Cheung16,Roy18,Roy21}. Local LL-AGN are associated to RIAFs \citep{Ho08LLAGN}, which have a tendency to drive powerful winds \citep{NarayanYi95,BlandfordBagel99} whose thermal and kinetic energy is deposited in the surrounding ISM. \cite{Almeida2023} developed a model for LL-AGN feedback through galaxy-scale winds produced by RIAFs and showed that these winds, especially if long-lived (on timescales of 10 Myr or longer) can prevent gas collapse and effectively quench star-formation in massive galaxies.There could be then a direct link between the LL-AGN activity at Cosmic Noon, which is responsible for rapid quenching, and the maintenance mode observed in local "red geysers".

%% Also note that the akcnowlodgment environment does not support long amounts of text. If you have a lot of people and institutions to acknowledge, do not use this command. Instead, create a new \section{Acknowledgments}.
\begin{acknowledgments}
We thank Salvatore Quai, Ivan Lopez, Raffaele Pascale, Marcella Brusa and Andrew Newman for 
the help and for the illuminating discussions.
We wish to thank the "Summer School for Astrostatistics in Crete" for providing training on the statistical methods adopted in this work. \\
The Blue Jay Survey is funded in part by STScI Grant JWST-GO-01810.
LB and SB are supported by the the ERC Starting Grant “Red Cardinal”, GA 101076080.
RD is supported by the Australian Research Council Centre of Excellence for All Sky Astrophysics in 3 Dimensions
 (ASTRO 3D), through project number CE170100013. RE acknowledges the support from grant numbers 21-atp21-0077, 
 NSF AST-1816420, and HST-GO-16173.001-A as well as the Institute for Theory and Computation at the Center for 
 Astrophysics. RW acknowledges funding of a Leibniz Junior Research Group (project number J131/2022).
This work is based on observations made with the NASA/ESA/CSA James Webb Space Telescope. The data were obtained 
from the Mikulski Archive for Space Telescopes at the Space Telescope Science Institute, which is operated by the
Association of Universities for Research in Astronomy, Inc., under NASA contract NAS 5-03127 for JWST. These 
observations are associated with program GO 1810. This work also makes use of observations taken by the 3D-HST 
Treasury Program (GO 12177 and 12328) with the NASA/ESA HST, which is operated by the Association of Universities 
for Research in Astronomy, Inc., under NASA contract NAS5-26555. 
\end{acknowledgments}


\vspace{5mm}
\facilities{JWST(NIRSpec)}


\software{\texttt{EMCEE} Python library \citep{emcee}, \texttt{Prospector} \citep{Prospector_article21},
    EA$z$Y \citep{eazycode}}



\appendix

\section{Full spectral measurements \label{appA}}
In Table~\ref{full_spec_measurements} we display the full set of spectral measurements for each galaxy in the sample. 
\begin{deluxetable}{cccccccccccc}
\rotate
%% Rotate to a landscape orientation
\tablewidth{1pt} 

\tablecaption{Spectral fitting results \label{full_spec_measurements}}

\tablehead{\colhead{COSMOS ID} & \colhead{$z_\mathrm{gas}$} & \colhead{$\sigma_\mathrm{gas}$} & \multicolumn{9}{c}{Flux} \\ 
\colhead{} & \colhead{}  & \colhead{($\rm
km~s^{-1}$)}& \multicolumn{9}{c}{($\rm
10^{-18}~erg~s^{-1}~cm^{-2}$)} \\
\cline{4-12} 
\colhead{} & \colhead{}  & \colhead{} & \colhead{[OII]} & \colhead{[NeIII]} & \colhead{H$\beta$} & \colhead{[OIII]5007}  & \colhead{H$\alpha$} & \colhead{[NII]6583} & \colhead{[SII]6716} & \colhead{[SII]6731} & \colhead{He I} 
}
\colnumbers
%% All data must appear between the \startdata and \enddata commands
\startdata
7549 & 2.6238 & $157.07_{-10.20}^{+10.40}$ & $2.63_{-0.26}^{+0.28}$ & $<0.020$ & $0.76_{-0.06}^{+0.06}$ & $<0.022$ & $2.14_{-0.13}^{+0.13}$ & $1.27_{-0.11}^{+0.12}$ & $0.68_{-0.11}^{+0.11}$ & $<0.651$ & $<0.201$ \\
8013 & 1.6894 & $261.08_{-3.94}^{+4.38}$ & $8.65_{-0.59}^{+0.60}$ & $1.06_{-0.14}^{+0.14}$ & $<0.987$ & $4.51_{-0.12}^{+0.13}$ & $9.13_{-0.13}^{+0.13}$ & $9.63_{-0.13}^{+0.14}$ & $3.29_{-0.29}^{+0.31}$ & $2.02_{-0.30}^{+0.32}$ & $1.59_{-0.16}^{+0.17}$ \\
8469 & \nodata & \nodata & $<0.126$ & $<0.212$ & $<0.643$ & $<3.75$ & $<0.142$ & $<0.098$ & $<0.129$ & $<0.100$ & $<0.201$ \\
9395 & 2.1268 & $400.47_{-26.93}^{+27.69}$ & $2.44_{-0.34}^{+0.35}$ & $<0.026$ & $<0.166$ & $<1.055$ & $<0.604$ & $<0.747$ & $<0.039$ & $<0.020$ & $<0.935$ \\
10128 & 1.8517 & $235.24_{-7.46}^{+8.01}$ & $0.55_{-0.03}^{+0.03}$ & $<0.002$ & \nodata & $0.95_{-0.04}^{+0.04}$ & $2.91_{-0.11}^{+0.11}$ & $2.93_{-0.11}^{+0.11}$ & $<1.649$ & $<0.873$ & $<0.261$ \\
10339 & 2.3636 & $536.04_{-54.90}^{+59.23}$ & $1.54_{-0.33}^{+0.34}$ & \nodata & $0.30_{-0.09}^{+0.10}$ & $<0.160$ & \nodata & \nodata & \nodata & \nodata & \nodata \\
10400 & \nodata & \nodata  & $<0.274$ & $<0.728$ & $<0.048$ & $<0.51$ & $<0.078$ & $<0.116$ & \nodata & \nodata & $<0.093$ \\
10565 & 2.4416 & $425.70_{-30.41}^{+31.50}$ & \nodata & \nodata & $<0.028$ & $<0.371$ & $0.29_{-0.20}^{+0.25}$ & $4.07_{-0.27}^{+0.26}$ & $<1.054$ & $<0.155$ & $3.12_{-0.27}^{+0.28}$ \\
10592 & \nodata & \nodata  & \nodata & \nodata & $<0.001$ & $<0.27$ & $<0.006$ & $<0.007$ & $<0.017$ & $<0.008$ & $<8.88$ \\
11142 & 2.4434 & $522.15_{-9.05}^{+9.26}$ & $6.58_{-0.24}^{+0.25}$ & $2.13_{-0.19}^{+0.19}$ & $<0.049$ & $17.18_{-0.31}^{+0.31}$ & $5.65_{-0.63}^{+0.62}$ & $35.42_{-0.63}^{+0.63}$ & $2.39_{-0.51}^{+0.52}$ & $4.53_{-0.51}^{+0.51}$ & $5.06_{-0.39}^{+0.38}$ \\
11494 & 2.0909 & $985.43_{-23.10}^{+10.85}$ & $7.59_{-1.77}^{+1.94}$ & $<0.167$ & $<0.123$ & $<0.060$ & $24.57_{-3.37}^{+2.76}$ & $2.68_{-1.91}^{+2.68}$ & $<0.702$ & $<0.512$ & $<21.30_{-1.89}^{+1.78}$ \\
16419 & 1.9257 & $437.70_{-10.40}^{+10.30}$ & $43.11_{-5.53}^{+5.56}$ & $<0.622$ & $<5.293$ & $14.97_{-0.62}^{+0.63}$ & $28.12_{-1.58}^{+1.61}$ & $25.03_{-1.45}^{+1.46}$ & $28.32_{-1.82}^{+1.87}$ & $18.04_{-1.86}^{+1.82}$ & $<1.610$ \\
18668 & 2.0857 & $204.44_{-3.43}^{+3.68}$ & $17.66_{-1.18}^{+1.15}$ & $<0.206$ & $3.74_{-0.21}^{+0.21}$ & $5.99_{-0.21}^{+0.21}$ & $29.28_{-0.45}^{+0.44}$ & $30.23_{-0.41}^{+0.42}$ & $6.84_{-0.35}^{+0.35}$ & $4.35_{-0.34}^{+0.34}$ & $4.44_{-0.32}^{+0.33}$ \\
%18688 & 2.0068 & $205.46_{-2.30}^{+2.31}$ & $16.01_{-0.44}^{+0.46}$ & $4.19_{-0.36}^{+0.36}$ & $9.06_{-0.32}^{+0.33}$ & $39.87_{-0.44}^{+0.44}$ & $48.60_{-0.72}^{+0.74}$ & $59.37_{-0.71}^{+0.74}$ & $7.48_{-0.63}^{+0.62}$ & $7.60_{-0.61}^{+0.61}$ & $7.57_{-0.55}^{+0.56}$ \\
%19572 & 1.8674 & $201.19_{-1.99}^{+2.06}$ & $9.98_{-1.05}^{+1.05}$ & $<2.360$ & $3.07_{-0.18}^{+0.17}$ & $25.62_{-0.22}^{+0.22}$ & $22.34_{-0.33}^{+0.31}$ & $42.17_{-0.31}^{+0.31}$ & $5.45_{-0.34}^{+0.33}$ & $4.23_{-0.31}^{+0.31}$ & $5.56_{-0.19}^{+0.19}$ \\
21477 & \nodata  & \nodata  & $<4.690$ & $<1.225$ & $<0.147$ & $<0.939$ & $<0.450$ & $<0.842$ & $<0.028$ & $<0.995$ & $<0.358$ \\
\enddata

%% General table comment marker
\tablecomments{Flux measurement for fitted emission lines in the quiescent sample.}

%% No \tablerefs indicated

\end{deluxetable}


\bibliography{biblio}
\bibliographystyle{aasjournal}



\end{document}